\documentclass[aps,pre,manuscript]{revtex4-1}
\usepackage{amsmath}
\usepackage{amssymb}
\usepackage{amsfonts}
\usepackage{amsthm}
\usepackage{graphicx}
\usepackage{subcaption}

\newcommand{\grad}{{\bf \nabla}}

\begin{document}
\author{Daria W. Atkinson}
\author{Christian D. Santangelo}
\affiliation{Department of Physics, University of Massachusetts Amherst}
\author{Gregory M. Grason}
\affiliation{Department of Polymer Science and Engineering, University of Massachusetts Amherst}

\begin{abstract}
Assemblies of one-dimensional filaments appear in a wide range of physical systems: from biopolymer bundles, columnar liquid crystals, and superconductor vortex arrays; to familiar macroscopic materials, like ropes, cables, and textiles.  Interactions between the constituent filaments in such systems are most sensitive to the {\it distance of closest approach} between the central curves which approximate their configuration, subjecting these distinct assemblies to common geometric constraints. In this paper, we consider two distinct notions of constant spacing in multi-filament packings in $\mathbb{R}^3$:   {\it equidistance}, where the distance of closest approach is constant along the length of filament pairs; and {\it isometry}, where the distances of closest approach between all neighboring filaments are constant and equal.  We show that, although any smooth curve in $\mathbb{R}^3$ permits one dimensional families of collinear equidistant curves belonging to a ruled surface, there are only two families of tangent fields with mutually equidistant integral curves in $\mathbb{R}^3$.  The relative shapes and configurations of curves in these families are highly constrained: they must be either (isometric) developable domains, which can bend, but not twist; or (non-isometric) constant-pitch helical bundles, which can twist, but not bend.  Thus, filament textures that are simultaneously bent and twisted, such as twisted toroids of condensed DNA plasmids or wire ropes, are doubly frustrated: twist frustrates constant neighbor spacing in the cross-section, while non-equidistance requires additional longitudinal variations of spacing along the filaments. To illustrate the  consequences of the failure of equidistance, we compare spacing in three ``almost equidistant''
ansatzes for twisted toroidal bundles and use our formulation of equidistance to construct upper bounds on the growth of longitudinal variations of spacing with bundle thickness.
\end{abstract}

\title{Constant spacing in filament bundles}

\maketitle

\section{Introduction}

Constant spacing between subunits governs a wide range of self-organized and manufactured pattern-forming assemblies \cite{aste_pursuit_2008}. At the smallest size scales, such assemblies arise generically as the ground states of a large family of interaction potentials.  Whether or not inter-element spacing is constant is fundamental to the behavior of materials, from the underlying processes of their formation, to their defects and distortions, and, ultimately, to their macroscopic responses (e.g. mechanical, optical).

The geometry of constant spacing and its implications for physical models of matter have been extensively studied for point-like (e.g. close-packings of spheres \cite{aste_pursuit_2008,conway_sphere_1998}) and surface-like (e.g. smectic liquid crystals \cite{kleman_points_1983}) subunits in three dimensions.  In comparison, the constant spacing of curve-like, quasi one-dimensional subunits, remains poorly understood.

\begin{figure*}[h!]
\begin{subfigure}[t]{.32\textwidth}
\includegraphics[width = \columnwidth]{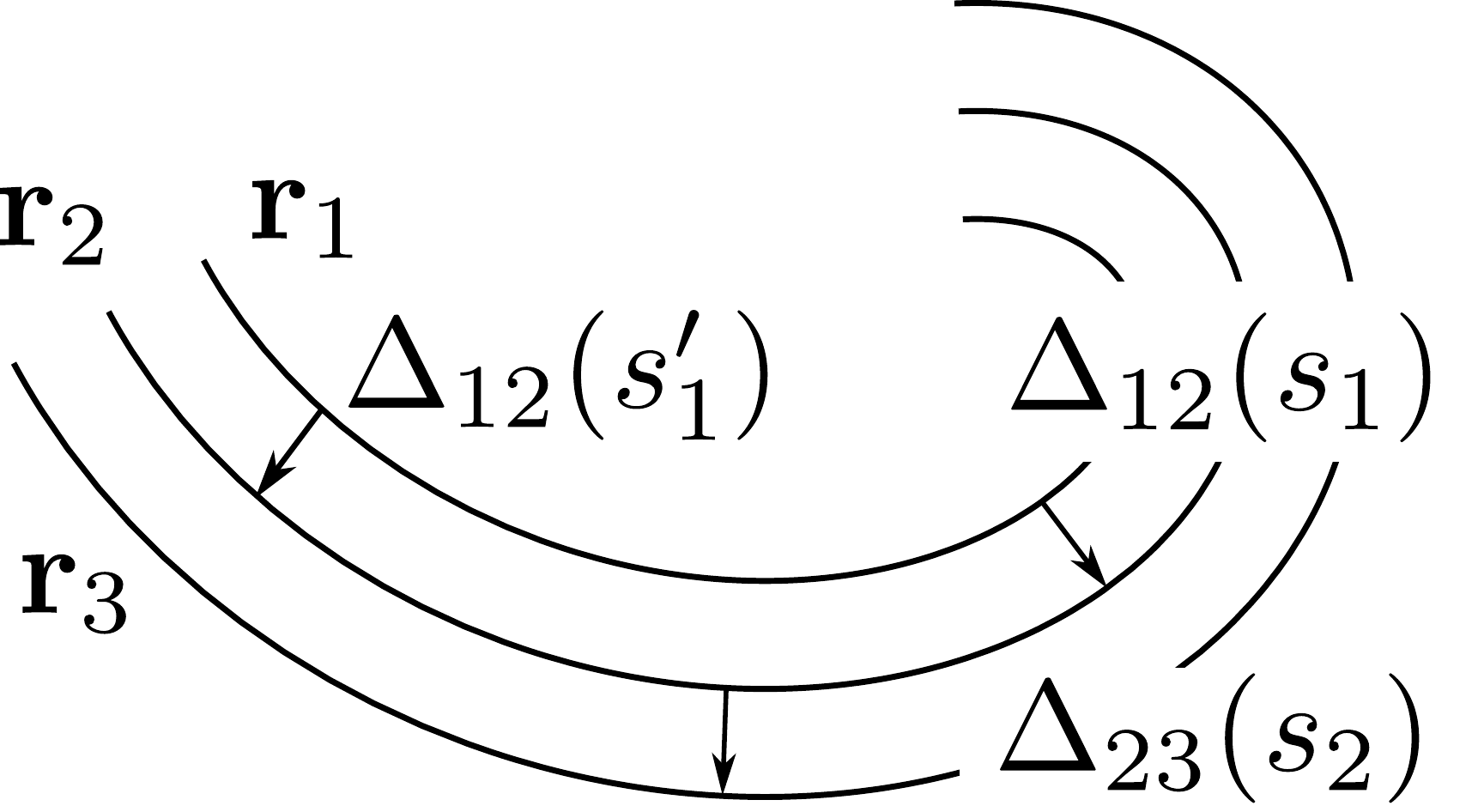}
\caption{}
\label{isometric}
\end{subfigure}
\begin{subfigure}[t]{.32\textwidth}
\includegraphics[width = \columnwidth]{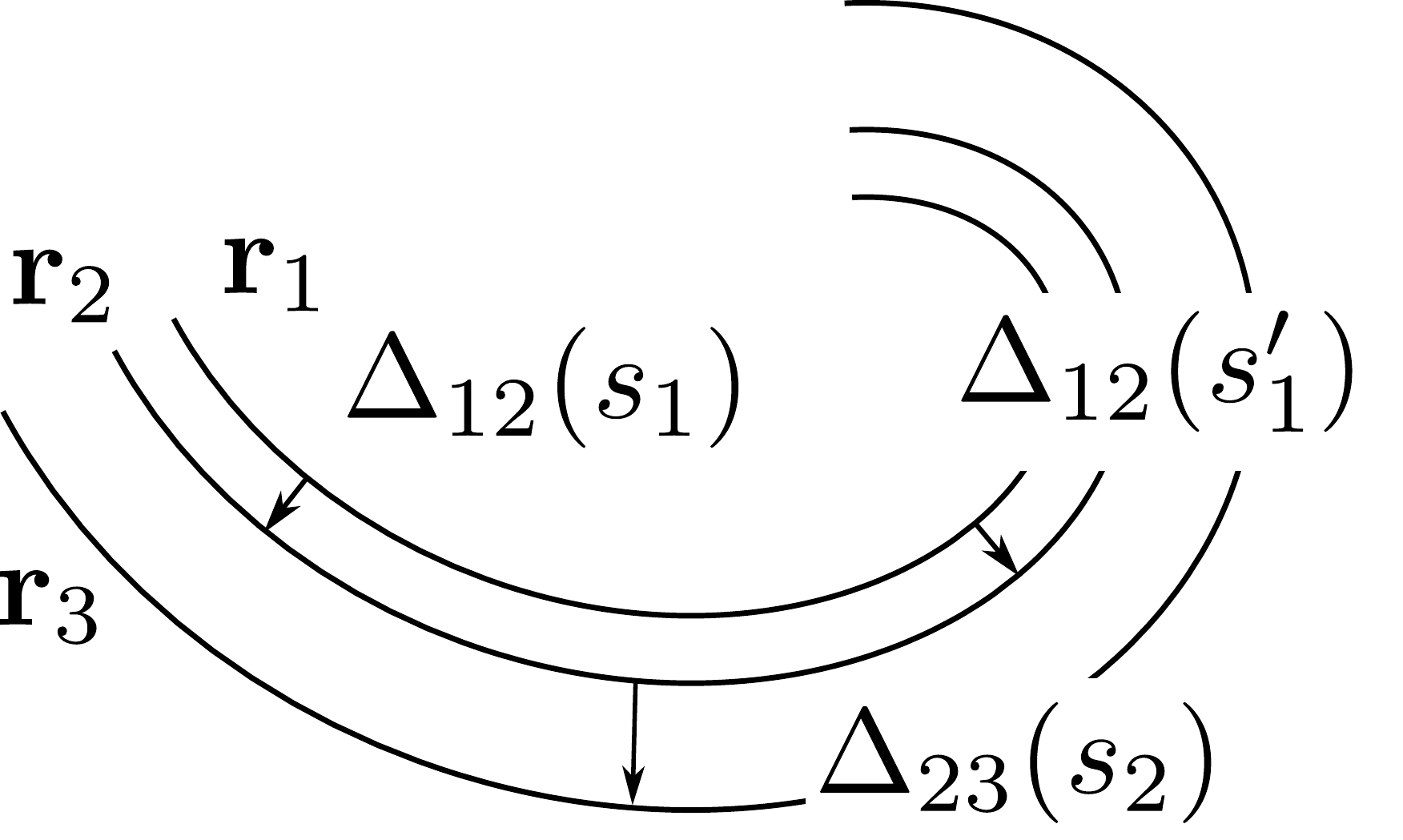}
\caption{}
\label{equidistance}
\end{subfigure}
\begin{subfigure}[t]{.32\textwidth}
\includegraphics[width = \columnwidth]{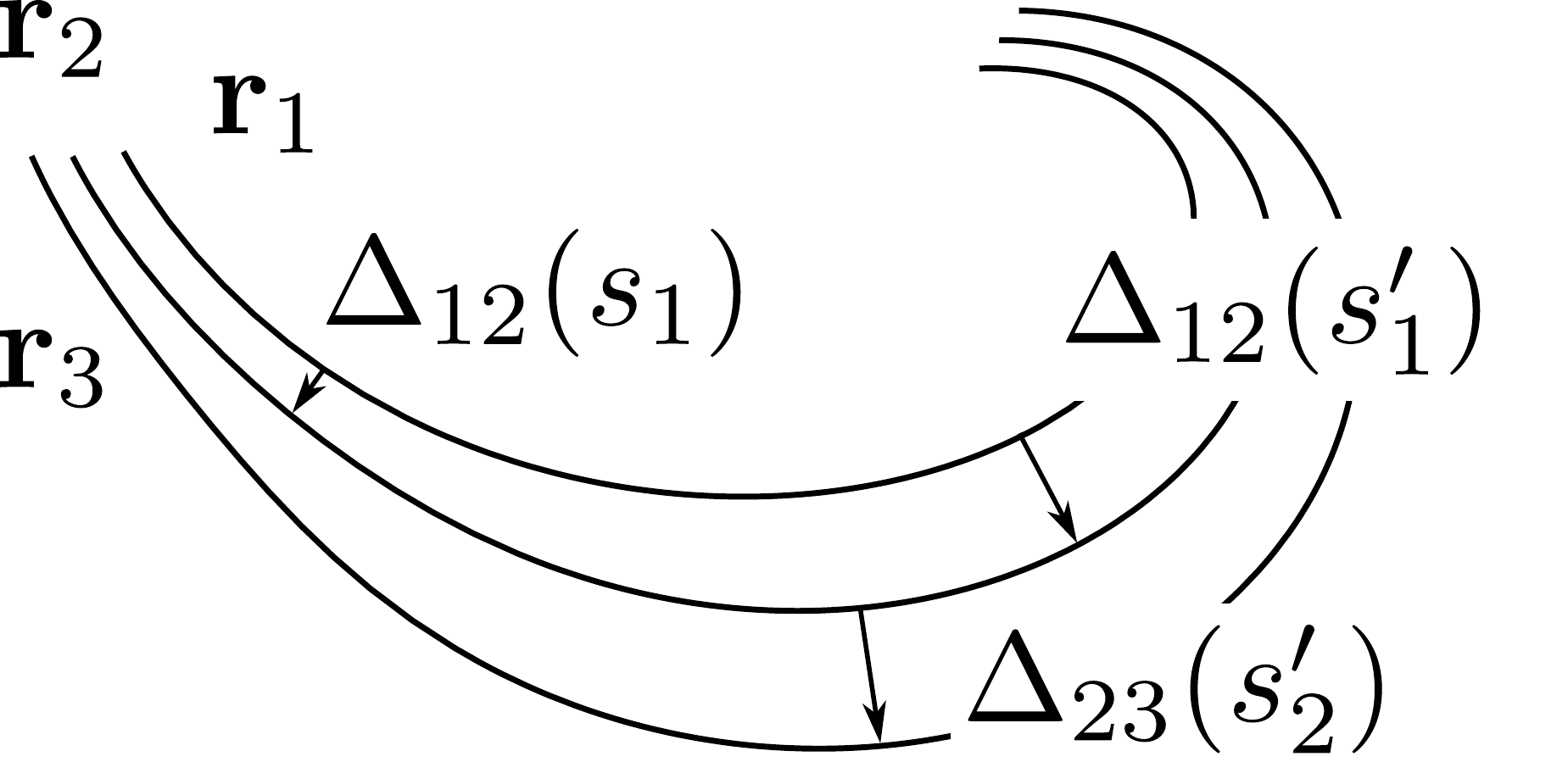}
\caption{}
\label{nonequidistance}
\end{subfigure}
\begin{subfigure}{.32\textwidth}
\includegraphics[height = .25 \textheight]{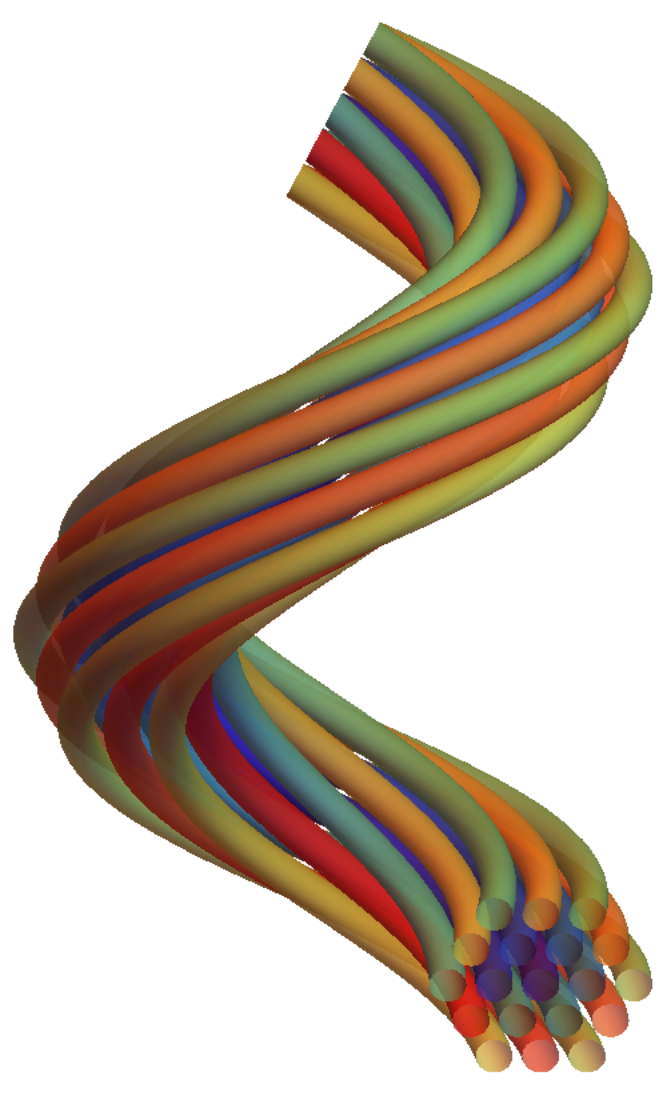}
\caption{}
\label{fig:devdom}
\end{subfigure}
\begin{subfigure}{.32\textwidth}
\includegraphics[height = .25 \textheight]{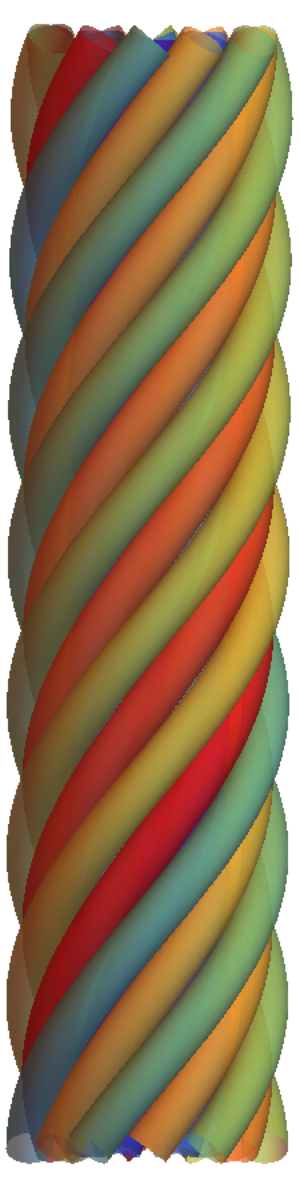}
\caption{}
\label{fig:helicaldom}
\end{subfigure}
\begin{subfigure}{.32\textwidth}
\includegraphics[height = .25 \textheight]{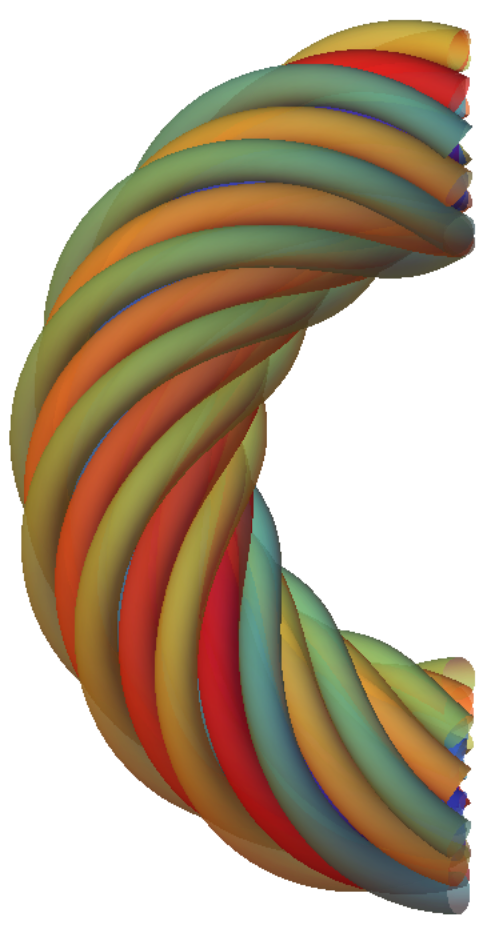}
\caption{}
\label{fig:toroid}
\end{subfigure}
\caption{Examples of curve arrays that illustrate the distinction between {\it equidistant} and {\it isometric} configurations.  The schematics in \ref{isometric}--\ref{nonequidistance} show three local inter-curve distances: $\Delta_{12}(s_1)$ and $\Delta_{12}(s_1')$, which denote the distance of closest approach between neighbor curves at arc positions $s_1$ and $s_1'$, respectively; and $\Delta_{23}(s_2)$, the distance of closest approach between an alternate pair.  In \ref{isometric}, an {\it equidistant} and {\it isometric} array (where $\Delta_{12}(s_1)=\Delta_{12}(s_1')=\Delta_{23}(s_2)$); in \ref{equidistance}, an {\it equidistant} but {\it non-isometric} array, ($\Delta_{12}(s_1)=\Delta_{12}(s_1')\neq \Delta_{23}(s_2)$); and in \ref{nonequidistance}, a {\it non-equidistant} array (where, in general,  
$\Delta_{12}(s_1)\neq \Delta_{12}(s_1')\neq \Delta_{23}(s_2)$).   While in two dimensions, every equidistant array is compatible with an isometric packing, there are equidistant, volume-filling curve textures of $\mathbb{R}^3$ which are incompatible with isometric packing \cite{bruss_non-euclidean_2012}.  As shown in Section~\ref{sec:fields}, there are only two families of equidistant curve fields in $\mathbb{R}^3$. Developable domains, as in \ref{fig:devdom}, are equidistant, and allow isometric filament packings \cite{bouligand_geometry_1980,kleman_developable_1980}, while helical domains, as in \ref{fig:helicaldom}, are equidistant, but do not allow isometric packings due to their effective positive Gaussian curvature \cite{bruss_non-euclidean_2012}.  Filament textures which are both bent and twisted, such as the toroidal bundle in \ref{fig:toroid}, cannot be equidistant.}
\label{equidistanceexplainer}
\end{figure*}

Perhaps the best studied regime of filament packings, motivated in part by physical models of protein and the packing of nucleic acids, arise from the close packing of a small number (typically, $N = 1$ or $2$) of plied or knotted flexible tubes \cite{banavar_colloquium_2003, banavar_self-interactions_2003, neukirch_geometry_2002, bozec_collagen_2007, bohr_close-packed_2011, snir_entropically_2005}.  In contrast, numerous physical scenarios -- from clumps of wet hair \cite{bico_adhesion_2004}, carbon nanotube yarns \cite{zhang_multifunctional_2004,thess_crystalline_1996} and biopolymer bundles \cite{livolant_highly_1989}  to macroscopic multi-filament wires and cables \cite{costello_theory_1990, gilbert_packing_1979} -- motivate the consideration of structures composed of an arbitrarily large number of filaments $N \gg 1$.   In 2D, the constraints on the constant spacing of $N \gg 1$ curves have been studied in the context of ordered stripe assemblies on variable shape surfaces \cite{santangelo_geometric_2007, knoppel_stripe_2015}.  Comparatively, packing $N \gg 1$ curves in a finite of volume of $\mathbb{R}^3$, which is most relevant to the structure of molecular fibers or macroscopic cables, introduces additional complexity due to two interrelated, but inequivalent notions of {\it constant spacing}.  In this paper, we call \emph{equidistant} families of curves for which the shortest distance between curves is constant along their length.  We then call {\it isometric} those equidistant families that permit uniform spacing between neighbors in their cross-section (see Figs.~\ref{isometric}--c).  At a pairwise level, equidistance is equivalent to constant surface contact between uniform diameter flexible tubes, and as such, is a natural way to describe optimal packings of cohesive filaments.

In this article, we present several results concerning the existence of families of equidistant curves in $\mathbb{R}^3$. We begin with a general introduction to ordered filament packings, outlining the differences between regular arrangements of filaments in two and three dimensions.  We show that, for any sufficiently smooth curve in $\mathbb{R}^3$, there exist families of non-parallel equidistant curves which cover a ruled surface, a natural generalization of the planar, parallel result. We then show that, for two such equidistant curves, it is always possible to place a third curve, which is equidistant to---but does not lie on the ruled surface spanned by---the first two curves.  Then, in order to understand the generic constraints of equidistance for $N\gg 1$ non-collinear curves,  we consider a continuum, vector field description of equidistant filament textures, which unlike the ruled surface families ``occupy" a finite 3D volume.  Solving explicitly for all unit-vector fields with sufficiently differentiable ($C^3$) equidistant integral curves, we show that equidistance imposes constraints on the first derivatives of the curves' tangents characterized by the vanishing of a two-component symmetric matrix, ${\bf H}$, of directional derivatives perpendicular to the local tangent.  Remarkably, and in stark contrast to the unconstrained equidistant triplets, there exist only two families of equidistant integral curves:  the {\it developable domains}, which can be bent, but not twisted \cite{kleman_developable_1980,starostin_perfect_2006}, (Fig.~\ref{fig:devdom}); and {\it helical domains} with constant pitch \cite{bruss_non-euclidean_2012}, which can be uniformly twisted, but not bent (Fig.~\ref{fig:helicaldom}).   We summarize the distinct features of these two families, outlining their compatibility with isometric packing and the constraints each family imposes on the relative shapes of curves in the packing.

In the remainder of the paper, we explore the consequences and limitations of this central result by numerically probing a simple family of ``almost equidistant'' filament bundles with both bend and twist: twisted toroidal bundles (Fig.~\ref{fig:toroid}). Such structures, are experimentally realized in systems of biopolymer condensates \cite{hud_cryoelectron_2001, leforestier_structure_2009, cooper_precipitation_1969}, and have recently gained interest as characterizing of a new class of topological soliton ``hopfion'' textures in liquid crystals \cite{ackerman_diversity_2017,ackerman_static_2017} and magnets \cite{sutcliffe_skyrmion_2017,sutcliffe_hopfions_2018,liu_binding_2018}.  We show that twisted toroids are a natural test bed for the structure of non-equidistant bundles, as the textures can continuously approach equidistance in the two asymptotic limits of {\it either} infinite major radius and finite twist (helical domain) {\it or} infinite helical pitch and finite curvature (developable domain). Because we expect the ground states of even complex, frustrated filament assemblies to minimize their deviations from uniform spacing, we approach this problem by comparing the growth of non-equidistance with twist and curvature using three ansatzes:  stereographic projections of the equidistant Seifert fibrations of $S^3$ into ${\mathbb R}^3$~\cite{kleman_frustration_1985, sadoc_3-sphere_2009}; splay-free tori, for which ${\rm Tr}({\bf H}) =0$~\cite{kulic_twist-bend_2004}; and a third class, characterized by ${\rm det}({\bf H}) =0$.  By constructing a numerical measure of non-equidistance, we compare asymptotic increases in non-equidistance with the lateral thickness (minor radius) of the twisted toroidal bundles, showing by construction that longitudinal variations between curves in the optimal structures will vanish at least as fast as thickness cubed in the limit of narrow bundles.

These results extend the understanding of geometric frustration in multi-filament packings well beyond previous studies, which have focused either on the frustration of filament and column shape in isometric packings \cite{bouligand_geometry_1980, kleman_developable_1980} or the frustration of the lateral spacing between filaments in non-isometric (twisted) packings \cite{bruss_non-euclidean_2012,grason_defects_2012,azadi_defects_2012,bruss_topological_2013}.  Specifically, this analysis highlights the nature of {\it longitudinal} frustration of constant spacing as distinct from, and complementary to, the {\it transverse} frustration of lateral spacing between neighbors in a large $N$ packing.  As experiments on isometric filament packings subject to twist have shown \cite{panaitescu_measuring_2017,panaitescu_persistence_2018}, the response of bundles to constraints of non-equidistance imposed by its global geometry will depend on the specifics of the filament packings.  Nevertheless, because the constraints for equidistance in these large $N$ packings are rather rigid, we anticipate several scenarios where the failure of equidistance triggers new structural and mechanical responses in physical models of bundles, including hierarchical packing of wires and cables.  

We conclude with a discussion of the bifurcation of equidistant bundles as additional curves are added,
conjecturing that there exists some finite $N_c>3$ such that any equidistant bundle with $N\geq N_c$ non-collinear curves falls into one of
the $N \gg 1$ families: either the helical or developable domains.

\section{Equidistance in multi-filament arrays}
In models of multi-filament packings, interactions between neighboring elements are often approximated by isotropic interactions between one-dimensional {\it central curves} \cite{cajamarca_geometry_2014, wang_twisting_2015, gonzalez_global_2002}.  In this context, local close-packing of two constant-diameter neighboring filaments requires that the distance of closest approach, $\Delta$, between their central curves is constant {\it along the entire length of the curves}.  In multi-filament bundles, uniform close-packing also requires that $\Delta$ is the same for any two nearest neighbors.   For simplicity, we call packings with longitudinally constant $\Delta$, as in Figs.~\ref{isometric} and \ref{equidistance}, {\it equidistant}, and those with uniform nearest neighbor distances, as in Fig.~\ref{isometric}, {\it isometric}.

Although equidistance is a necessary condition for isometric packing, it is useful to consider the implications of equidistance independent of isometry.  Equidistant packings are particularly valuable as they reduce the problem of inter-element distances in a three-dimensional bundle to the lower dimensional problem of packing elements on a two-dimensional surface.  This perspective has enabled in-depth explorations of the (non-isometric) ground-state structure of close-packed, twisted bundles \cite{bruss_non-euclidean_2012, hall_morphology_2016}.  Beyond this, cohesive interactions naturally impose a cost for variations in the local spacing between attractive filaments, and it is therefore natural to anticipate that equidistant geometries (if they are compatible with topological constraints or mechanical loading) are ground-state configurations of many models, particularly when inter-filament cohesion dominates over the mechanical costs of intra-filament bend and twist.

At a pairwise level, the conditions for equidistance are found by demanding that the shortest distance between two curves, $\mathbf{r}_1$ and $\mathbf{r}_2$, is constant along their arc lengths, $s_1$ and $s_2$, respectively.  This is shown by considering the closest separation from ${\bf r}_1$ at $s_1$ to ${\bf r}_2$, which can be defined as $\Delta_{12}(s_1) \equiv {\rm min}_{s_2}\big[ | {\bf r}_1(s_1)-{\bf r}_2(s_2)| \big]$.  For a given $s_1$, this requires that the closest arc position, $s_2=s_2(s_1)$ on ${\bf r}_2$, satisfies
\begin{equation}
\Big( \partial_{s_2}  \big|\mathbf{r}_1(s_1) - \mathbf{r}_2(s_2)\big|^2\Big)_{s_2=s_2(s_1)}  = -2 ~ {\bf T}_2 \big[s_2(s_1) \big] \cdot {\bf \Delta}_{12} (s_1) = 0,
\label{closestapproacheqn}
\end{equation}
where ${\bf T}_2 = {\bf r}_2'\big[s_2(s_1)\big]$ is the tangent to ${\bf r}_2$ at the distance of closest approach, and ${\bf \Delta}_{12} (s_1) = {\bf r}_1(s_1)-{\bf r}_2 \big[s_2(s_1) \big]$ is the closest separation vector to ${\bf r}_2$ from ${\bf r}_1(s_1)$~\footnote{In general, there are may be multiple extrema of $\big|\mathbf{r}_1(s_1) - \mathbf{r}_2(s_2)\big|^2$, corresponding to multiple solutions for $s_2(s_1)$ to Eq. (\ref{closestapproacheqn}), for a given pair.  Our analysis assumes the minimal distance for solutions $s_2(s_1)$ for a given $s_1$.}.   The solution to this condition induces a reparameterization $s_2(s_1)$ of $\mathbf{r}_2$ in terms of $s_1$, such that we can rewrite this second curve as ${\bf r}_2(s_1) \equiv {\bf r}_2\big[ s_2(s_1) \big]$.  Equidistance between $\mathbf{r}_1$ and $\mathbf{r}_2$ then requires that
$\Delta_{12}(s_1)$ is constant in $s_1$, so
\begin{equation}
\partial_{s_1} \big|{\bf \Delta}_{12} (s_1) \big|^2 = 2~ \big[{\bf T}_1(s_1)-\frac{\partial s_2(s_1)}{\partial s_1}{\bf T}_2(s_1) \big] \cdot {\bf \Delta}_{12} (s_1) =  0.
\label{equidistance_eqn}
\end{equation}
While Eq.~\eqref{equidistance_eqn} is generically quite difficult to solve explicitly, when $s_2(s_1)$ is invertible ($ \partial s_2/  \partial s_1 \neq 0$), it has a straightforward geometric interpretation.  In particular, ${\bf \Delta}_{12}(s_1)$ has constant magnitude, and remains perpendicular to the tangents of both $\mathbf{r}_1$ and $\mathbf{r}_2$ at the points of closest approach, $s_1$ and $s_2(s_1)$, respectively.  In the language of, e.g., Ref.~\cite{cantarella_minimum_2002}, equidistant curves pairs are {\it doubly-critical} at all points.

\subsection{Equidistance in the Plane}
\label{subsec:plane}
For plane curves, as shown in Figs.~\ref{isometric}--c, a pair of curves $\mathbf{r}_1$ and $\mathbf{r}_2$ can be written in terms of the local distance between the two curves, $\Delta_{12}$, the arc length $s_1$ of ${\bf r}_1$, $\mathbf{r}_1$, and its normal, ${\bf N}_1$, as
\begin{equation}
\mathbf{r}_2 (s_1) = {\bf r}_1(s_1) +\Delta_{12}(s_1) {\bf N}_1(s_1).
\end{equation}
If the two filaments are equidistant (i.e. $\partial_{s_1} \Delta_{12} = 0$), then the curves must be parallel (i.e. $ {\bf T}_1 = {\bf T}_2$) at the points of closest approach.  It is then straightforward to embed a field of curves ${\bf r}_n$ that are all parallel to ${\bf r}_1$, using a similar parameterization $\mathbf{r}_n (s_1)= {\bf r}_1(s_1) + \Delta_{n}  {\bf N}_1(s_1)$, where $ \Delta_{n} $ is the distance between the $n$th curve and ${\bf r}_1$.  Note that $\Delta_{n}$ can be extended only up to the global radius of curvature of ${\bf r}_1$, at which point $\mathbf{r}_n$ becomes singular and its distance map from ${\bf r}_1$ becomes noninvertible \cite{gonzalez_global_1999}.  If $\Delta_{n+1}-\Delta_{n}$ is constant for all $n$, then the equidistant curves are also {\it isometric}.

Hence, any planar curve ${\bf r}_1(s_1)$ can be extended to an equidistant family on the plane (at least in a neighborhood of $ {\bf r}_1(s_1)$ smaller than its global radius of curvature), and every equidistant family is compatible with isometric packing.  As there are no constraints imposed by constant spacing on the shape of ${\bf r}_1(s_1)$ (beyond smoothness), we say that packings of planar curves are {\it unfrustrated}.

\subsection{Equidistant pairs and ruled surfaces in $\mathbb{R}^3$}
\label{subsec:pairsandsurfaces}
\begin{figure}
\begin{subfigure}{.32\textwidth}
\includegraphics[width=\columnwidth]{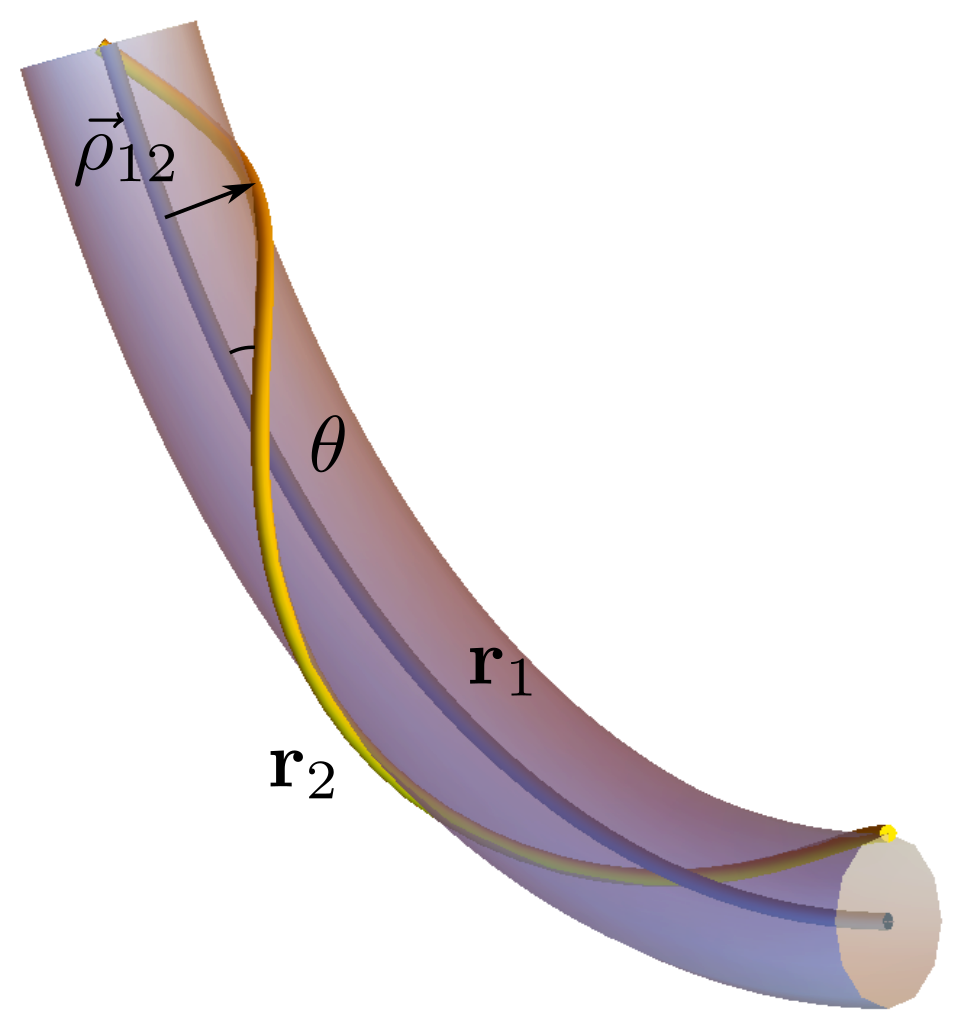}
\caption{}
\label{fig:1tube}
\end{subfigure}
\begin{subfigure}{.32\textwidth}
\includegraphics[width=\columnwidth]{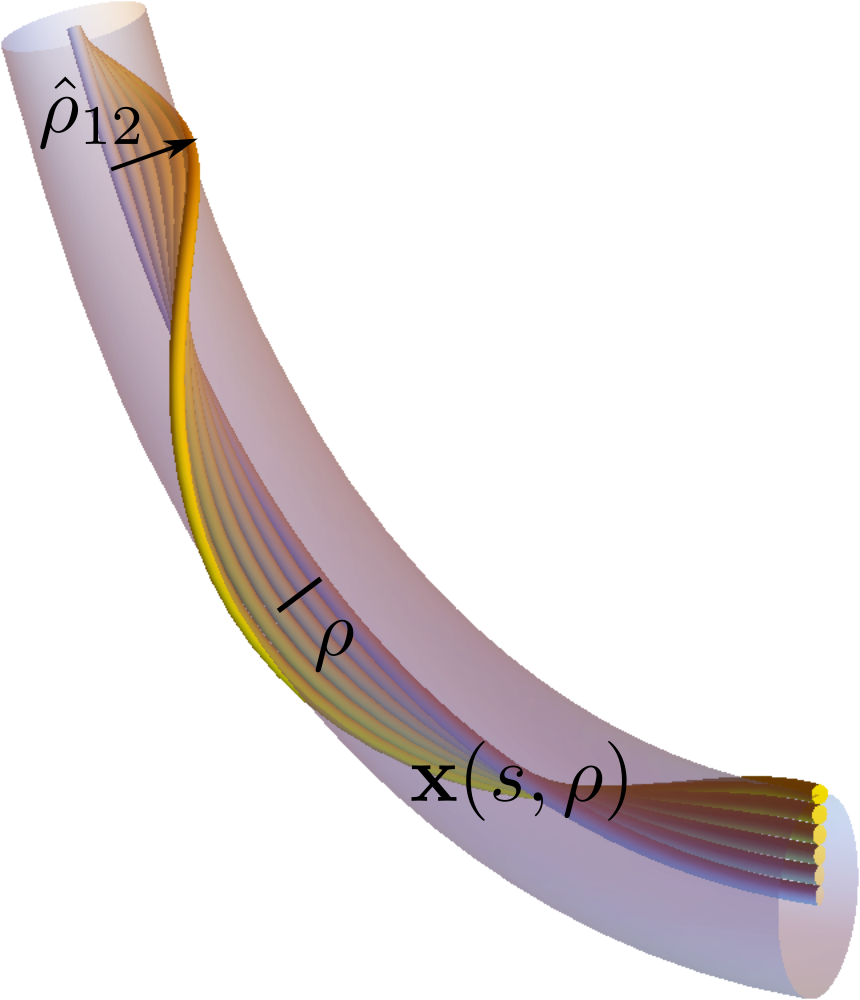}
\caption{}
\label{fig:ruledsurface}
\end{subfigure}
\begin{subfigure}{.32\textwidth}
\includegraphics[width=\columnwidth]{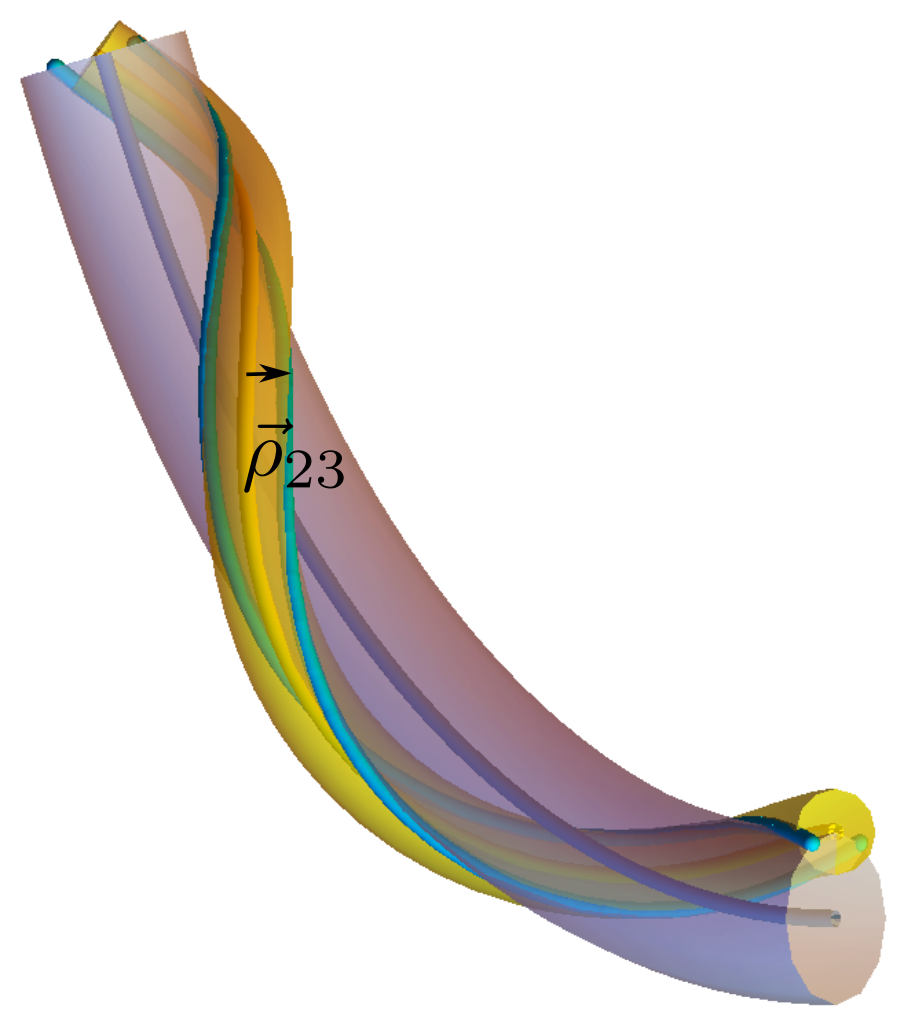}
\caption{}
\label{fig:2tube}
\end{subfigure}
\caption{A simple heuristic argument suggests that for any sufficiently smooth space curve $\mathbf{r}_1$ in $\mathbb{R}^3$ there are at least
two additional curves, $\mathbf{r}_2$ and $\mathbf{r}_3$ such that $\mathbf{r}_1$, $\mathbf{r}_2$, and $\mathbf{r}_3$ are all
equidistant.  To see this, imagine extending a tube of constant radius $\rho_{12}$ around $\mathbf{r}_1$, as in \ref{fig:1tube}.
A curve $\mathbf{r}_2$ that lives on this tube is equidistant to $\mathbf{r}_1$.  This construction can be extended to a one-dimensional family of equidistant and isometric curves, as in \ref{fig:ruledsurface}, where the curves mark lines of constant $\rho$ on the ruled {\it separating surface} generated by $\mathbf{r}_1$ and $\mathbf{r}_2$.  These filaments are equidistant, as shown in Eq.~\eqref{ruledsurface}, and isometric,
as the distance of closest approach between neighboring curves is equivalent for any pair of neighbors. Extending a tube of radius $\rho_{23}$ (not necessarily equal to $\rho_{12}$) around
$\mathbf{r}_2$, as in \ref{fig:2tube}, we see that the curve $\mathbf{r}_3$, which traces out the intersection of the tubes, is equidistant to both
$\mathbf{r}_1$ and $\mathbf{r}_2$, but in contrast to the separating surface in \ref{fig:ruledsurface}, the three curves are {\it not} collinear.}
\label{fig:tubeconstruction}
\end{figure}

In contrast, the geometry of equidistant pairs of curves in $\mathbb{R}^3$ is much more flexible than that of planar curves. For a curve $\mathbf{r}_1$ in three dimensions, there are two linearly independent directions locally perpendicular to ${\bf T}_1$. Notably, this means that there are curves $\mathbf{r}_1$ and $\mathbf{r}_2$ that are equidistant but {\it not} parallel, so that ${\bf T}_1 \neq {\bf T}_2$ at the points of closest approach (i.e. points separated by ${\bf \Delta}_{12}(s_1)=-{\bf \Delta}_{21}(s_2)$).
Furthermore, as we show in Appendix~\ref{appdx:existence}, for any sufficiently differentiable curve $\mathbf{r}_1$ and
distance $\Delta_{12}$ less than the global radius of curvature, there exist multiple curves $\mathbf{r}_2$ such
that $\mathbf{r}_1$ and $\mathbf{r}_2$ are equidistant but not parallel.  Heuristically, one can understand this flexibility in terms of the ``tubular" construction illustrated in Fig.~\ref{fig:1tube}, where a circular tube of fixed radius $\Delta_{12}$ encloses ${\bf r}_1$.  Any curve, ${\bf r}_2$, on this tubular surface for which ${\bf T}_1 \cdot {\bf T}_2=\cos \theta_{12}$ has a constant sign is equidistant to ${\bf r}_1$.

Given any two equidistant curves $\mathbf{r}_1$ and $\mathbf{r}_2$, there is an infinite family of equidistant curves that lie along a ruled surface spanned by the vectors, ${\bf \Delta}_{12}(s)= \mathbf{r}_2(s) - \mathbf{r}_1(s)$, which we call the {\it separating surface} \footnote{We again adopt the reparameterization of $\mathbf{r}_2$ in terms of the arc length of $\mathbf{r}_1$, which we call $s$ for simplicity of notation.}. To see this, let $\hat{\rho}_{12}(s) \equiv (\mathbf{r}_2(s) - \mathbf{r}_1(s))/\Delta_{12}$.  Then, we define a family of curves, parameterized by the distance $\rho$ from ${\bf r}_1$ towards ${\bf r}_2$,
\begin{equation}
\mathbf{r}_\rho(s) = \mathbf{r}_1(s) + \rho~\hat{\rho}_{12}(s).
\label{rhohat}
\end{equation}
It is straightforward to verify the equidistance of two curves at $\rho_1$ and $\rho_2$ by verifying that their tangents are perpendicular to their separation vector. Specifically,
\begin{equation}
\partial_s \mathbf{r}_{\rho_1}(s) \cdot \big[\mathbf{r}_{\rho_1}(s) - \mathbf{r}_{\rho_2}(s')\big] = \Big(\mathbf{T}_1(s) + \frac{\rho_1}{\Delta_{12}}\Big[\frac{ \partial s_2}{\partial s} \mathbf{T}_2(s) - \mathbf{T}_1(s)\Big] \Big) \cdot \big[\mathbf{r}_{\rho_1}(s) - \mathbf{r}_{\rho_2}(s')\big],
\end{equation}
which is zero when $s' = s$ because curves $\mathbf{r}_1$ and $\mathbf{r}_2$ are equidistant with distance of closest approach at $s$.  The equivalent necessary condition for $\mathbf{r}_{\rho_2}$ also holds.   This family of equidistant curves forms a ruled surface, the separating surface of $\mathbf{r}_1$ and $\mathbf{r}_2$,
\begin{equation}
\mathbf{x}_{12} (s,\rho) = \mathbf{r}_1(s) + \rho~\hat{\rho}_{12}(s),
\label{ruledsurface}
\end{equation}
ruled by the vectors $\hat{\rho}_{12}(s)$ (as shown by Fig. \ref{fig:ruledsurface}). 

The regular spacing of curves on one such surface, the helicoid, has been suggested by Archad, et. al.~\cite{achard_liquid_2005} as an explanation for the structure of the $B7^*$ phase of bent core liquid crystals \cite{coleman_polarization-modulated_2003}.  These ruled separating surfaces are also a natural generalization  of the equidistant plane curves discussed in Subsection~\ref{subsec:plane} to three dimensions, showing that the torsion of one or both curves allows for equidistant curves to be non-parallel.  Any sufficiently smooth curve in $\mathbb{R}^3$ permits such  ruled surface families, and, as in the planar case, a subset of the equidistant curves on a separating surface can always be chosen such that the curves are isometric.

\subsection{Non-collinear equidistant triplets}
\label{subsec:triplets}

The families of equidistant and isometric curve packings described above are strictly two-dimensional, as they lie on the ruled, separating surface that is uniquely defined for any equidistant pair in $\mathbb{R}^3$.  Before continuing on to the problem of three-dimensional fields of equidistant curves, we first give a simple construction to show that it is generically possible, for a given equidistant pair, ${\bf r}_1$ and ${\bf r}_2$, to find at least one additional curve, ${\bf r}_3$, which is mutually equidistant to the first two, but that does not lie on their separating surface. 

As shown in Fig. \ref{fig:2tube}, we can illustrate the constraints of equidistance by surrounding the curves with tubes of fixed radii perpendicular to their local tangents.  This guarantees that the separation vector between the curves has constant length (say, $\Delta_{12}$), is along the radial direction, and is, by construction, perpendicular to the central curve (say, ${\bf r}_1$) and the curve defined on its surface (say, ${\bf r}_2$).  Likewise, it is straightforward to construct tubes around the two equidistant curves ${\bf r}_1$ and ${\bf r}_2$.  The radii for these tubes can be chosen rather arbitrarily (up to the limits placed by the global radius of curvature) to be $\Delta_{13}$ and $\Delta_{23}$. These tubes intersect along two curves that do not lie on the ruled surface spanning $\mathbf{r}_1$ and $\mathbf{r}_2$, but are, by construction, equidistant to both of those curves.  Either one these curves can be taken as ${\bf r}_3$, forming an equidistant triplet.

We note that while the geometry of three equidistant, non-collinear curves constructed sequentially, as described above, is relatively flexible, it is far from clear how the addition of more curves alters the constraints on their shapes and relative arrangement.  For example, adding a fourth equidistant curve to the triplet in Fig. \ref{fig:2tube}, requires the intersection of {\it three} tubular surfaces surrounding those curves along a single 1D curve, a condition that can only be satisfied for a subset of equidistant triplets, a point we return to in the discussion.

\section{Fields of Equidistant Curves}
\label{sec:fields}

Thus motivated to find families of multi-curve packings corresponding to bundles of $N \gg 1$ non-collinear filaments in $\mathbb{R}^3$, we adopt a continuum description based on the integral curves of unit vector fields.  
In many physical examples of bundles, like DNA condensates or carbon nanotube ropes, a combination of dense-packing and intra-filament stiffness keeps filaments in quasi-parallel orientation.  In such a dense, multi-filament bundle (in the absence of filament ends in the array), the geometry of a finite set of backbone curves indexed by $m$, $\{\mathbf{r}_m(s)\}$, can be analyzed by a unit vector field $\mathbf{t}({\bf x})$ that smoothly interpolates between their tangents, so that $\mathbf{t}\big[\mathbf{r}_m(s)\big] = \hat{t}_m(s)$.

\begin{figure}
\includegraphics[width = .6\columnwidth]{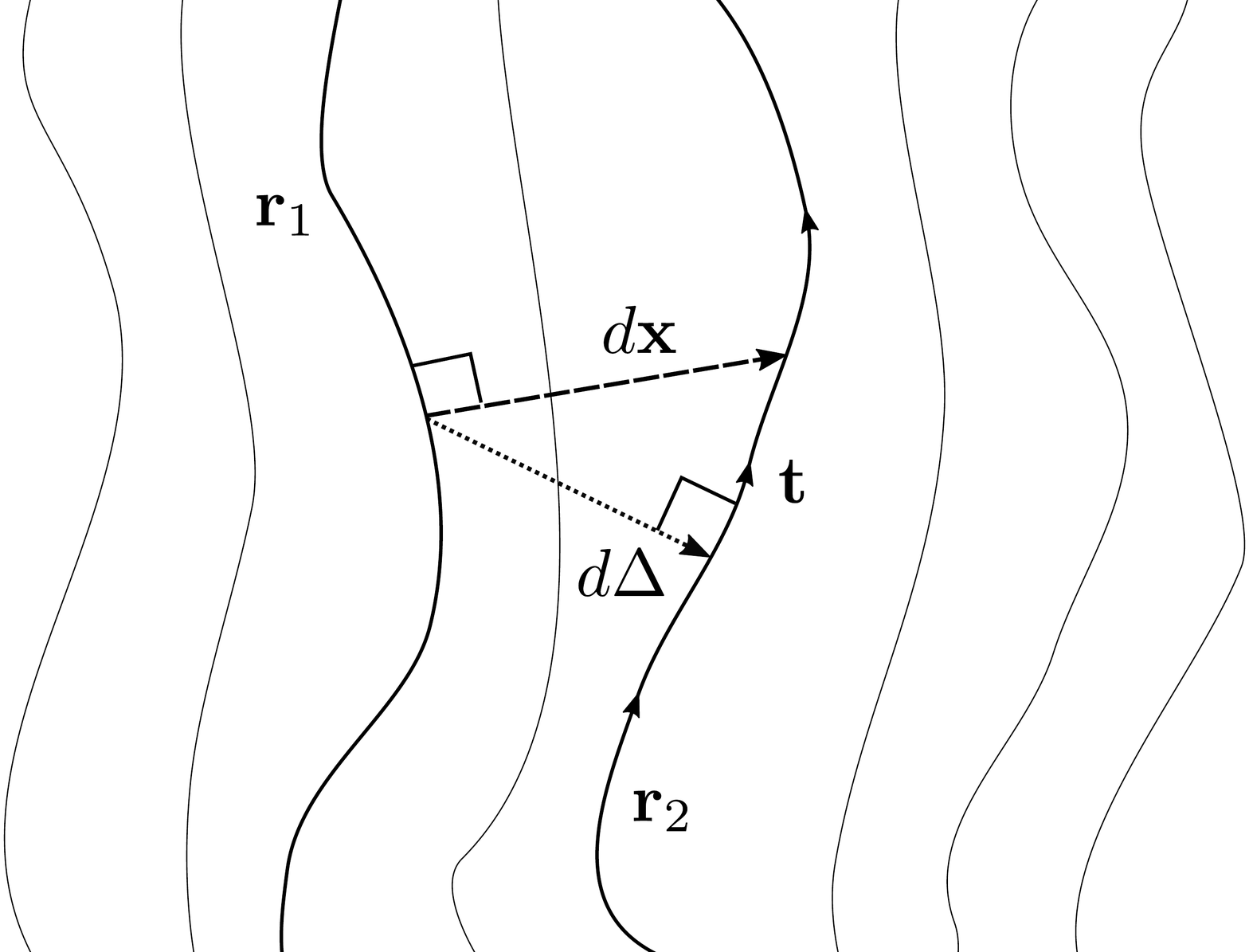}
\caption{The distance of closest approach $\Delta$ from a curve $\mathbf{r}_1$ to a curve $\mathbf{r}_2$ is perpendicular to the tangent $\hat{t}$
of $\mathbf{r}_2$.  Given a tangent field $\mathbf{t}$, the distance of closest approach between $\mathbf{r}_1$ at $\mathbf{x}$ and
$\mathbf{r}_2$ at $\mathbf{x}+ d\mathbf{x}$ can be found by projecting out the tangent field $\mathbf{t}$, giving an infinitesimal distance
of closest approach $d\Delta = d\mathbf{x} - \mathbf{t} (\mathbf{t} \cdot d\mathbf{x})$.}
\label{closestapproach}
\end{figure}

In this section, we derive the conditions under which the integral curves of a given unit tangent field $\mathbf{t}({\bf x})$ are all mutually equidistant in a region of $\mathbb{R}^3$. These families of fields of equidistant curves are particularly valuable for physical models of multi-filament bundles, in that they permit the embedding of an arbitrary number of equidistant curves in a finite volume of three-dimensional space, in contrast to the 2D submanifolds of $\mathbb{R}^3$ spanned by ruled separating surfaces.   In the following section, we show that conditions imposed by equidistance lead to strong constraints on the relative shapes and orientations of the integral curves in the set.
\subsection{Local metric and convective flow tensor}
\label{convectiveflow}
Given a unit tangent field $\mathbf{t}:\mathbb{R}^3 \mapsto S^2$, we can find the distance of closest approach between two integral curves that pass through infinitesimally close points $\mathbf{x}$ and $\mathbf{x} + d\mathbf{x}$ by projecting out the component of $d\mathbf{x}$ along $\mathbf{t}$, as shown in Figure \ref{closestapproach}.  The resulting local distance of closest approach is given by
\begin{equation}
d\Delta^2 = (\delta_{ij} - t_i t_j) dx^i dx^j.
\label{metricdef}
\end{equation}
We note that this projection can be written as a 2D metric 
\begin{equation}
g_{ij} ({\bf x}) = \delta_{ij} - t_i({\bf x}) t_j({\bf x})
\label{eq: localmetric}
\end{equation}
by considering $d{\bf x}$ in a planar section of $\mathbb{R}^3$ whose normal ${\bf N}$ satisfies ${\bf N} \cdot \mathbf{t}(\mathbf{x})>0$ in some region (e.g. a plane which is perpendicular to $\mathbf{t}(\mathbf{x})$ at some ${\bf x}$) \cite{bruss_non-euclidean_2012,grason_colloquium_2015}.

In this local formulation, the distance between two curves is constant along their length when $\partial_s d\Delta^2 = 0$, where $\partial_s={\bf t} \cdot \nabla$ is the directional derivative along ${\bf t}$.  Differentiating, and using the convective flow of the separation between integral curves $\partial_s d{\bf x} = d{\bf x} \cdot \nabla \mathbf{t}({\bf x})$, we find that
\begin{equation}
\partial_s d\Delta^2 = [\partial_i t_j + \partial_j t_i -  t_k \partial^k (t_i t_j + t_j t_i)]dx^i dx^j=h_{ij} dx^i dx^j.
\label{hijref}
\end{equation}
Because $\mathbf{t}$ is a unit vector, and hence $t_i \partial_k t_i=0$, $h_{ij}$ is zero for all components along $\mathbf{t}$.  The remaining terms belong to a 2D block whose components can be associated with locally orthonormal directions $\hat{\bf e}_1 ({\bf x})$ and $\hat{\bf e}_2 ({\bf x})$ that span the plane perpendicular to ${\bf t}({\bf x})$ (i.e., $\hat{\bf e}_1 ({\bf x}) \times \hat{\bf e}_2({\bf x})={\bf t}({\bf x})$).  Projecting $h_{ij}$ onto this two-dimensional basis defines
\begin{equation}
H_{\alpha \beta} \equiv (\hat{\bf e}_\alpha)_i  h_{ij}  (\hat{\bf e}_\beta)_j
\end{equation}
where $\alpha, \beta =1,2$.  $H_{\alpha \beta}$ is a symmetric, 2-tensor, which we call the {\it convective flow tensor}, that measures the longitudinal deviations from equidistance.  Hence, equidistance requires $H_{\alpha \beta} =0$.

These conditions can be recast in terms of the directional derivatives of the tangent field perpendicular to ${\bf t}({\bf x})$, 
\begin{equation}
(\nabla {\bf t})^{(2D)}_{\alpha \beta} \equiv  (\hat{\bf e}_\alpha)_i  \partial_i t_j  (\hat{\bf e}_\alpha)_j
\end{equation}
from which we have $H_{\alpha \beta} = (\nabla {\bf t})^{(2D)}_{\alpha \beta} + (\nabla {\bf t})^{(2D)}_{\beta \alpha}$.  Therefore, a field $\mathbf{t}$ is equidistant only when these transverse directional derivatives are skew symmetric, with
\begin{equation}
(\nabla {\bf t})^{(2D)}_{\alpha \beta} =  f({\bf x})  \epsilon_{\alpha \beta}\ \ \ \ \ {\rm for \ } H_{\alpha \beta} =0 ,
\label{eq: skew}
\end{equation}
where $f$ is any function and $\epsilon_{\alpha \beta}$ is the totally antisymmetric Levi-Civita symbol.  This skew-symmetric structure is closely related to the {\it double-twist texture} of the blue phases of chiral liquid crystals \cite{wright_crystalline_1989}.  In the context of the blue phases, it is well appreciated that the geometry of $\mathbb{R}^3$ is incompatible with uniformly double-twisted textures \cite{sethna_relieving_1983}, leading to the formation of defect-ordered phases of finite-diameter double-twist tubes.  In the context of the present problem, however, the condition of Eq.~\eqref{eq: skew} is slightly weaker, and the rate of double-twist, as parameterized by the function $f({\bf x})$, may vary spatially without disrupting the equidistance of the field lines.  

Before moving on to solve for the equidistant curve fields, we note that the equidistance of integral curve fields promotes the metric description of Eq.~\eqref{eq: localmetric} from one that measures local distances between infinitesimally spaced curves, to one in which the metric $g_{ij}({\bf x})$ relates the true Euclidean distances of closest approach of {\it finitely-separated curves} to their coordinate separations in some reference plane (e.g. in a given 2D plane cutting through ${\bf t}({\bf x})$).  When $H_{\alpha \beta}({\bf x})=0$ everywhere within some volume, distances of closest approach between finitely separated curves can be found as geodesic arc lengths computed according to the induced metric.  In the language of differential geometry, equidistance is the necessary and sufficient condition for a Riemannian foliation, where the metric properties of the leaves (curves) inherited from the embedding space (the distance of closest approach in $\mathbb{R}^3$) are encoded by the Riemannian metric of a lower dimensional base manifold (in this case, a 2D surface) \cite{gromoll_metric_2009}.  In the following section, we classify the isometry of equidistant curve fields in terms of the Gaussian curvature of these foliations.

\subsection{Equidistant solutions}
\label{equisol}

\begin{figure}
\includegraphics[width = .5\columnwidth]{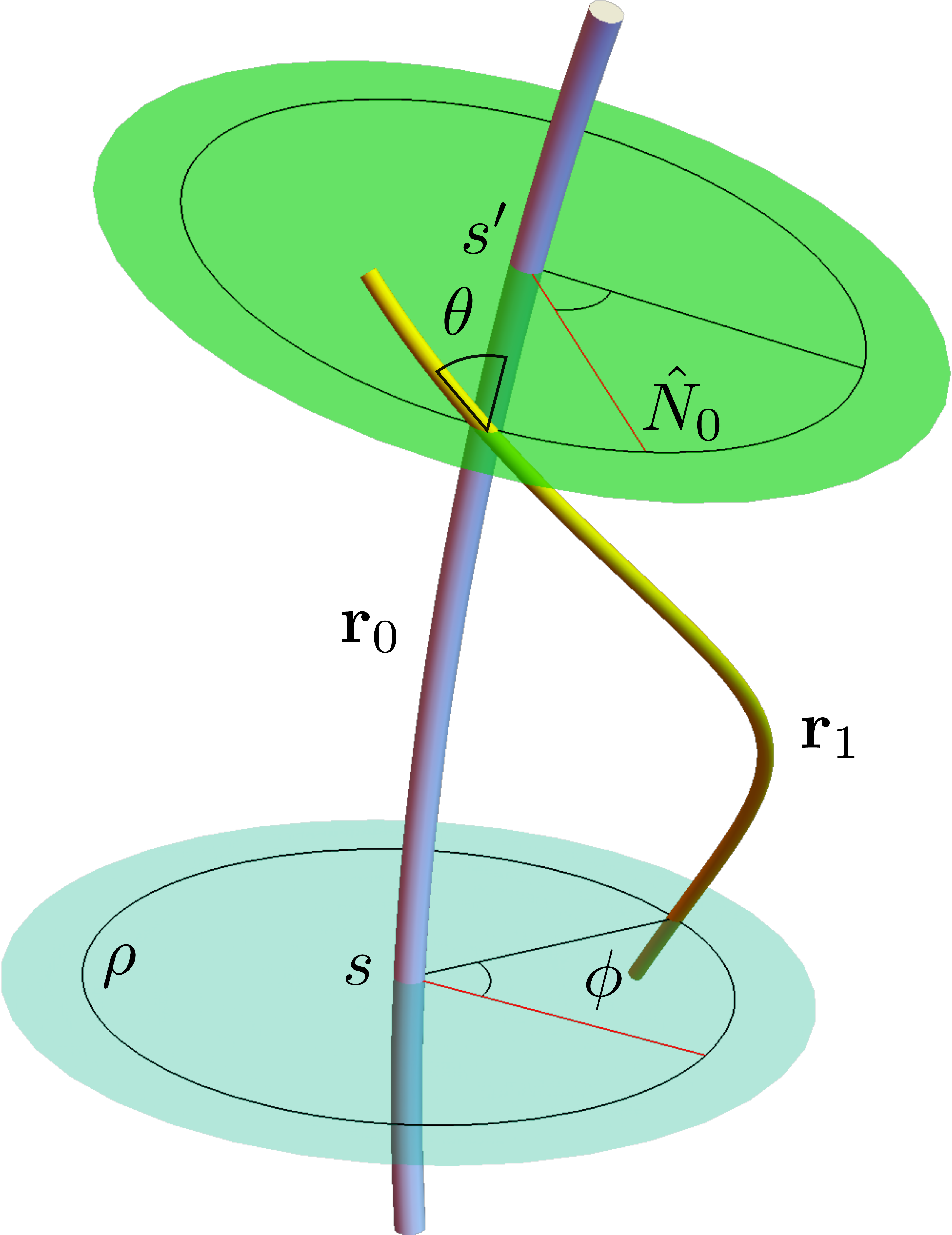}
\caption{A schematic diagram of the coordinates described in Eq.~\eqref{coorddef}, showing the radial distance $\rho$ from a central curve,
$\mathbf{r}_0$; $\phi$, the polar angle in the plane normal to $\mathbf{r}_0$ measured with respect to the principle normal, $\hat{N}$; and $s$, the position along $\mathbf{r}_0$ in terms of its arc length.}
\label{fig:coords}
\end{figure}

The skew symmetry of $(\nabla \mathbf{t})^{(2D)}$ in Eq. \eqref{eq: skew} gives three independent differential equations for $\mathbf{t}$, which can be solved to find every equidistant tangent field.  We begin by choosing coordinates $\{s,\rho, \phi\}$ adapted to some integral curve $\mathbf{r}_0$ of the tangent field, where $s$ is an arc length parameterization of $\mathbf{r}_0$, $\rho$ is a polar distance in the plane perpendicular to $\hat{t}_0$ at some $s$, and $\phi$ the polar angle in the same plane (see Appendix~\ref{appdx:coords} for details), such that
\begin{equation}
\mathbf{x}(s,\rho,\phi) = \mathbf{r}_0(s) + \rho ~\hat{\rho}(s,\phi),
\label{coorddef}
\end{equation}
as shown schematically in Fig \ref{fig:coords}. In these coordinates, any field $\mathbf{t}$ whose integral curves are equidistant to $\mathbf{r}_0$ will be perpendicular to the separation vector $ \rho ~\hat{\rho}(s,\phi)$ and hence can be written
\begin{equation}
\mathbf{t}(s,\rho,\phi) = \cos{\big[\theta(s,\rho,\phi)\big]} \hat{t}_{0}(s) + \sin{\big[\theta(s,\rho,\phi)\big]} \hat{\phi}(s,\phi),
\label{eq: tform}
\end{equation}
where $\hat{t}_{0} \equiv \partial_s \mathbf{r}_0$, $\hat{\phi} \equiv \partial_\phi \hat{\rho}$ and
$\theta$ is a scalar field which characterizes the tilt of integral curves with respect to $\hat{t}_0$.

We can analyze the components of $(\nabla \mathbf{t})^{(2D)}_{\alpha \beta}$ in the two orthonormal directions, $\hat{\rho}$ and $\hat{b} = {\bf t} \times \hat{\rho}$, in the plane normal to $\mathbf{t}$ at ${\bf x}$.  All $\mathbf{t}({\bf x})$ of this form satisfy $(\nabla \mathbf{t})^{(2D)}_{\rho \rho} =0$ explicitly.  Using the coordinate transformations given in Appendix ~\ref{appdx:coords}, the other components of $\nabla \mathbf{t}$ can be found exactly:
\begin{equation}
(\nabla \mathbf{t})^{(2D)} = \begin{bmatrix} 0 & \partial_\rho \theta \\
- \frac{\sin{\theta} \cos{\theta}}{\rho(1 - \rho \kappa_0 \cos{\phi})} & \frac{1}{\rho}
\partial_\phi \theta \big(\frac{\sin{\theta} \tau_0 \rho}{1 - \rho \kappa_0 \cos{\phi}} + \cos{\theta} \big)
+ \frac{\sin{\theta}(\kappa_0 \sin{\phi} - \partial_s \theta)}{1-\rho \kappa_0 \cos{\phi}} \end{bmatrix} .
\label{gradttensor}
\end{equation}
The skew symmetry of $(\nabla \mathbf{t})^{(2D)} $ required for equidistance gives us the differential equations:
\begin{align}
(H_{\rho b} = 0) & \ \ \ \ \partial_\rho \theta = \frac{\sin{\theta}\cos{\theta}}{\rho (1-\rho \kappa_0 \cos{\phi})} \label{termsandconditions1}  \\
(H_{b b} = 0) &\ \ \ \  \sin{\theta} \partial_s \theta = [(\tfrac{1}{\rho} - \kappa_0 \cos{\phi})\cos{\theta} +
\tau_0\sin{\theta}] \partial_\phi \theta + \kappa_0 \sin{\phi}\sin{\theta}  .
\label{termsandconditions2}
\end{align}
The first of these differential equations, Eq.~\eqref{termsandconditions1}, can be integrated directly, giving us
\begin{equation}
\tan{ \theta} = \Omega(s,\phi)\frac{\rho}{1-\rho \kappa_0 \cos{\phi}},
\label{genericrhosolution}
\end{equation}
where $\Omega(s,\phi)$ is a constant of $\rho$.  Substituting into Eq. \eqref{termsandconditions2} and rearranging, we find that:
\begin{multline}
H_{b b} =0 =- \big[(1-\rho \kappa_0 \cos{\phi})^2 + (\rho \Omega)^2\big]^{-3/2}
\bigg\{\big[ \rho^2 \tau_0 \Omega(1-\rho \kappa_0 \cos{\phi}) 
+ (1 - \rho \kappa_0 \cos{\phi})^3 \big]\partial_\phi \Omega \\ - \rho^3 \Omega^2 \partial_s \kappa_0 \cos{\phi}  
+ \big[\rho \kappa_0 \Omega \cos{\phi} - \Omega\big ] \rho^2 \partial_s \Omega+
\rho^3 \kappa_0 \Omega^2(\tau_0 - \Omega)\sin{\phi}\bigg\},
\label{eq:Hbb}
\end{multline}
where $\partial_s {\bf t}_0 = \kappa_0 {\bf n}_0$ and $\partial_s ( {\bf t}_0 \times {\bf n}_0) = -\tau_0 {\bf n}_0$ give, respectively, the curvature ($\kappa_0$) and torsion ($\tau_0$) of the reference curve ${\bf r}_0$.   The numerator of Eq.~\eqref{eq:Hbb} is a cubic polynomial of $\rho$, so, grouping by powers of $\rho$ and recognizing that solutions to $H_{b b} =0$ require the coefficients of these linearly independent terms to vanish, we find only two possible solutions for equidistant fields.  In the first case we have
\begin{equation}
\Omega = 0,
\label{developable}
\end{equation}
which gives us solutions that are locally parallel in the plane normal to $t_0$ (i.e., $\theta =0$).  The second family of solutions require
\begin{align}
\partial_\phi \Omega &=  0  \nonumber \\
\partial_s \Omega& = 0  \nonumber \\
\kappa_0' \cos{\phi} &= \kappa_0(\Omega - \tau_0) \sin{\phi},
\label{helical}
\end{align}
so that every twisted equidistant field has constant pitch $\Omega$, and includes an integral curve with constant curvature and torsion.  Because any curve with constant curvature and torsion is a helix, the torsion is fixed by the pitch $\Omega$, and the curvature is fixed by the torsion and its distance from some straight line, this second family of solutions is the one parameter family of bundles of constant pitch circular helices.  

\section{The Equidistant Packings}
\label{sec:packings}
In the previous section, we find that the conditions for equidistance are only satisfied by two restrictive families of curve fields, corresponding to the respective conditions in Eqs. \eqref{developable} and \eqref{helical}.  In this section, we describe in turn the geometric properties of these two families and the physical scenarios in which they have been invoked.   We focus on the distinguishing features of inter-filament texture, intra-filament shape, inter-filament spacing (or metric geometry), and constraints on the lateral thickness of bundles of smoothly embeddable curves.  

Motivated by applications of multi-filament packing in liquid crystals and soft matter \cite{kulic_twist-bend_2004,rey_liquid_2010,shin_filling_2011,brown_equilibrium_2014}, it is natural to analyze the inter-filament texture in terms of the Frank elastic gradients of the tangent field, in particular, first derivatives of  ${\bf t}$ that constitute generalized ``orientational strains'' in the Frank-Oseen free energy \cite{degennes_physics_1995}.  Because $(\grad \cdot {\bf t}) = \text{tr}(H)$, all equidistant curve fields are splay-free.  The twist, $ {\bf t} \cdot ( \grad \times {\bf t})$,
provides a measure the neighbor-average inter-filament skew angle in the packing, that is, the local rate of mutual rotation of neighbors~\cite{grason_braided_2009}.  The final first-order Frank term is associated with bending of the tangent field, that is, it is a measure of {\it intra-filament curvature} $\kappa$, which is computed from the convective derivative of ${\bf t}$ itself, namely $({\bf t} \cdot \grad )  {\bf t}= \kappa {\bf n}$ where again, ${\bf n}({\bf x})$ is the local normal to the integral curve at ${\bf x}$.  In addition to the curvature, intra-filament shape is characterized by the {\it torsion} $\tau$ which is given by the rotation of the binormal ${\bf b} = {\bf t} \times {\bf n}$ around the tangent, $({\bf t} \cdot \grad )  {\bf b} = - \tau {\bf n}$ .

In addition to these measures of intra- and inter-filament gradients we analyze the metric properties of the equidistant packings in terms of the Gaussian curvature $K$ of the 2D metric $g_{ij} ({\bf x})$ induced on a planar section through the bundle, as in Eq.~\eqref{eq: localmetric}, which may be directly derived via standard formulas \cite{millman_elements_1977}.  Finally, we define the {\it maximum thickness} as the diameter of a bundle of filaments that can be smoothly extended normal to a given central curve in the packing.  That is, beyond this maximum thickness, continuing the equidistant field introduces shape singularities in the integral curves, features which we exclude from our analysis due to the prohibitive costs of kinks in physical realization of multi-filament packings.  Table \ref{tab: equidistant} summarizes the geometric comparisons between the two families of equidistant curve fields.  We describe each family in turn.

\begin{table*}
\begin{tabular}{|l|c|c|c|c|c|}
\hline
{\bf Equidistant family} & {\bf Twist} & {\bf Curvature} & {\bf Torsion} & {\bf Metric curvature} & {\bf Max thickness} \\ \hline
\hline
Developable domains & 0 & $\frac{\kappa_0}{1-\kappa_0 \rho \cos \phi }$ & $\frac{\tau_0}{1-\kappa_0 \rho \cos \phi }$ & 0 & $\min_s[1/\kappa_0]$ \\ \hline
Helical domains (const. $\Omega$) & $\frac{2 \Omega}{ 1 + (\Omega \rho)^2  }$ &  $\frac{ \Omega^2 \rho}{ 1 + (\Omega \rho)^2  }$ &  $\frac{ \Omega}{ 1 + (\Omega \rho)^2  }$  & $\frac{3 \Omega^2}{[ 1+ (\Omega \rho)^2 ]^2 }$ & $\infty$ \\ \hline
\end{tabular}
\caption{Summary of geometric properties of the distinct families of equidistant curves.  Inter-filament {\it twist} is defined by ${\bf t} \cdot (\nabla \times {\bf t} )$.  {\it Curvature}, $\kappa$, of filaments at ${\bf x}$ is derived from $({\bf t} \cdot \nabla) {\bf t} = \kappa {\bf n}$, while {\it torsion}, $\tau$, is derived from  $({\bf t} \cdot \nabla) {\bf b} = -\tau {\bf n}$, where ${\bf n}$ and ${\bf b} = {\bf t} \times  {\bf n}$ are the normal and binormal, respectively.  The {\it metric curvature} is the Riemannian curvature of the inter-filament metric $g_{ij} ({\bf x})$, and the {\it max thickness} describes the largest lateral diameter of the domain that is embeddable without self intersection.  For developable domains, generalized cylindrical coordinates are given with respect to a reference curved of respective curvature and torsion, $\kappa_0$ and $\tau_0$, and for helical domains, coordinates are defined with respect to a straight central curve.}
\label{tab: equidistant}
\end{table*}

\subsection{($\Omega = 0$): Developable Domains}
The first equidistant family, described by Eq.~\eqref{developable},  corresponds to what have been called {\it developable domains} (see example in Fig.~\ref{fig:devdom}).  These textures were originally described by Bouligand \cite{bouligand_geometry_1980} and fully classified by Kl\'{e}man \cite{kleman_developable_1980} in the context of columnar liquid crystals.  Developable domains have neither twist (i.e. ${\bf t} \cdot (\grad \times {\bf t} ) =0$) nor splay ($\grad \cdot {\bf t} = 0$), and thus the filament tangents are all parallel at the point of closest approach, their tangents are normal to a common set of planes (i.e. $\theta =0$), and the closest separations between curves lie in these 2D planes.  Hence, it is straightforward to see that their metric geometry is Euclidean.  Indeed, the developable domains are the {\it only isometric family} of $N \gg 1$ curves in $\mathbb{R}^3$.

Because the curves are normal to a common set of planes and they do not twist around one another, they also share the same Frenet frames at points of closest contact, giving closely related shapes.  Constructing a developable domain around a given curve with curvature $\kappa_0$ and torsion $\tau_0$, the shape of all other curves in the domain are fully determined \cite{starostin_perfect_2006}, such that
\begin{equation}
\kappa({\bf x}) = \frac{ \kappa_0}{1-\kappa_0 \rho \cos \phi}; \ \tau({\bf x}) = \frac{ \tau_0}{1-\kappa_0 \rho \cos \phi},
\end{equation}
where $\rho$ is the closest distance to the central curve and $\phi$ is the angle between the separation to the reference curve and its normal (see Fig.~\ref{fig:coords}).  Hence, for non-zero bending, these normal planes intersect along the cuspidal edge of the developable surface generated by the locus of all the centers of curvature of the filaments in the bundle~\cite{bouligand_geometry_1980}.    Bouligand and Kl\'{e}man argued that such curvature singularities manifest as characteristic topological defects in columnar phases.  Here, we argue further that this same geometry places constraints on the maximum size of isometric filament packings with finite bending. While the developable domains permit isometric filament packings and can be embedded around reference curves of any (smooth) shape, embeddings of finite curvature filaments are spatially limited to a thickness around the central curve less than its global curvature radius \cite{gonzalez_global_1999} as they become singular along this developable surface.

\subsection{($\Omega \neq 0$): Constant-Pitch, Helical Domains}
We first discuss the second equidistant family, described by Eq.~\eqref{helical}, in terms of a straight central curve (i.e. $\kappa_0 = 0$) that threads through its center along an axis $\rho =0$  (see example in Fig.~\ref{fig:helicaldom}).  Relative to this axis, these curves are easily seen to be helices with a tilt angle, $\theta = \arctan (\Omega \rho)$, with respect to the center which increases with radius $\rho$, but has constant pitch $2 \pi /\Omega$ (the corresponding curvature and torsion are given in Table~\ref{tab: equidistant}).   Indeed, the geometry of these equidistant {\it helical domains} closely corresponds to the ``double-twist tube'' that is the fundamental building block of the liquid crystal blue phases \cite{wright_crystalline_1989}.    Unlike the developable domains, which do not permit twist, this second family is twisted, with ${\bf t} \cdot (\nabla \times {\bf t} ) = 2 \Omega/[1 + (\Omega \rho)^2 ]$.  As inter-filament twist is generically favored in chiral filamentous materials such as biopolymer assemblies \cite{neville_biology_1993,bouligand_liquid_2008,grason_braided_2009}, helical domains are important structural models of the compromise between the preference for chiral inter-filament packing and the cohesive preference for equidistance.  Recent experiments show further that the constant-pitch helical texture emerges in mechanically twisted filament packings \cite{panaitescu_measuring_2017,panaitescu_persistence_2018}.

While helical domains are the only twisted family of equidistant curves in $\mathbb{R}^3$, twist is incompatible with isometric packing in the cross section \cite{bruss_non-euclidean_2012,grason_colloquium_2015}.  This can be seen from the metric in polar coordinates (as defined in Fig.~\ref{fig:coords}) centered on the straight curve:
\begin{equation}
g =
\begin{bmatrix}
1 & 0 \\
0 & \rho^2 \cos^2{\theta}.
\end{bmatrix}
\end{equation}
Because $\cos{\theta}=1/\sqrt{1+(\Omega \rho)^2}$ decreases with $\rho$, hoops of constant distance from the center are effectively shortened relative to the Euclidean plane, consistent with positive Gaussian curvature \cite{gromoll_metric_2009},
\begin{equation}
K = \frac{3 \Omega^2}{(1+ \Omega^2 \rho^2)^2}.
\end{equation}
The effect of this positive Gaussian curvature is to frustrate constant lateral spacing of filaments (e.g. equi-triangular packing).  Physical models of twisted cohesive bundles have shown that this metric frustration promotes accumulation of inter-filament stresses \cite{hall_how_2017} or else stabilize topological defects \cite{bruss_non-euclidean_2012} in the cross sectional order of twisted cohesive bundles.  
Notably, the Gaussian curvature of helical domains is concentrated in the core, as the metric flattens in the limit $\Omega \rho \to \infty$.  Hence, the disruption of uniform lateral spacing at the core of helical domains notwithstanding, this equidistant family can be extended smoothly to fill all of $\mathbb{R}^3$, in contrast to the spatially limited, developable domains.

While the above description assumes a straight central curve, the choice of the central curve is arbitrary, provided that it satisfies Eqs.~\eqref{helical}, such that it is a helix whose torsion is equal to $\Omega$.  It is straightforward to show that choosing one such helix simply gives a reparameterization of the same family of helical domains.  For example, in terms of generalized cylindrical coordinates $(\rho',\phi')$ around a reference curve with curvature $\kappa_0$ we have the Gaussian curvature distribution,
\begin{equation}
K = \frac{3 \Omega^2}{\big[(1 - \rho' \kappa_0 \cos{\phi'})^2 + (\Omega \rho')^2\big]^2}.
\end{equation}
It can be shown that this metric derives from considering a planar slice through the helical bundle that is normal to a curve at finite radius, $\kappa/(\kappa^2+\Omega^2)$.

Thus, up to the orientation and position of a central axis of rotation, every equidistant helical domain is parameterized by a single real number, $\Omega$, which can be viewed as a simple rescaling of the same structure.

\section{Almost Equidistant Bundles}
In the previous section, we showed that equidistant curve packings fall into two strict families.  These two families are either strictly untwisted but arbitrarily bent, or uniformly twisted around a straight axis.  In this section, we illustrate the consequences of falling outside these strict geometrical constraints for inter-filament spacing in multi-filament bundles (e.g. a bundle that is simultaneously bent {\it and} twisted). Such generic geometric conditions are encountered in widely varying rope-like structures, from hierarchical strands of wire-ropes \cite{costello_theory_1990}, to twisted, curved bundles of condensed biopolymers~\cite{cooper_precipitation_1969, hud_cryoelectron_2001, leforestier_structure_2009}.  

Here, we study arguably the simplest possible non-equidistant geometry, the {\it twisted toroidal bundles}, a family of architectures that conveniently spans both equidistant families (see example, Fig.~\ref{fig:toroid}).  Notably, several previous models of close-packed toroidal bundles have been developed to describe the structure and thermodynamics of biopolymer toroids.  A primary focus of many of these model has been the relationship between their geometry and their orientational order~ \cite{kulic_twist-bend_2004, koning_saddle-splay_2014} without regard to their metric geometry.  Work of Sadoc, Charvolin and others have considered idealized metric geometries possible in $S^3$, but to date, the limits to the uniformity of filament spacing in toroids embedded in $\mathbb{R}^3$ have not been explored.

Below we consider three ansatzes for non-equidistant, twisted-toroidal bundles.  Two are related to previous models of either ``splay-free'' bundles or projections of ideal fibrations of $S^3$ to Euclidean space.  In the context of the present study, we can contrast all three ansatzes in terms of the structure of the convective flow tensor ${\bf H}$.  As described in Sec.~\ref{convectiveflow}, ${\bf H}$ describes the first-derivative of the local separation between integral curves and equidistance requires all three independent components of $H_{\alpha \beta}$ to vanish.  Forcing a bundle to be simultaneously bent and twisted hence requires at least one of the components to be non-zero.  Below, we compare the variable filament spacing in three toroidal ansatzes: stereographic projection of the Seifert fibrations of $S^3$ to $\mathbb{R}^3$, for which $H_{\alpha \beta}= H({\bf x}) \delta_{\alpha \beta}$; splay-free toroidal bundles, for which ${\rm Tr}[{\bf H}]=0$; and twisted toroidal bundles, for which ${\rm det}[{\bf H}]=0$.  

To compare the inter-filament spacing within these toroidal ansatzes quantitatively, we construct bundles from integral curves of each construction.  The cross section of each bundle has 1+6+12 filaments, whose initial centers are chosen from three concentric layers of a hexagonal packing of unit spacing.  Each filament is then discretized in to $N=10000$ arc positions, from which the distance matrix between all positions on each neighboring filament pair is calculated.    Minimizing over the set of distances between a point $s_i$ on curve $\mathbf{r}_i$ and all the positions $s_j$ in $\mathbf{r}_j$, gives the  distance of closest approach from $\mathbf{r}_i(s_i)$ to $\mathbf{r}_j$, $\Delta_{ij} (s_i)$.  Averaging over positions $s_i$ gives the average separation from $i$ to $j$, $\langle \Delta_{ij} \rangle$.  To compare longitudinal uniformity of inter-filament spacing in these distinct textures, we define the following measure of local deviation from equidistance:
\begin{equation}
\delta r_i(s_i) =\frac{1}{n_{i}} \sum_{\langle ij \rangle}  \frac{ \Delta_{ij} (s_i)-\langle \Delta_{ij} \rangle}{\langle \Delta_{ij} \rangle},
\end{equation}
where $\sum_{\langle ij \rangle}$ denotes the sum over the $n_i$ neighbors of the $i$th filament in the initial hexagonal packing.  The Supplemental Video shows an example of the variation of $\delta r_i(s_i)$ throughout a bent and twisted packing (generated via the $det({\bf H})0$ ansatz described below).

This quantity measures the extent to which a point $s_i$ on $\mathbf{r}_i$ is relatively closer or further than its average separation from other filaments in the bundle.  We define a measure of the total variability of spacing in the bundle $\langle \delta r^2 \rangle$ as the average of the square of this local measure over the lengths of all filaments,
\begin{equation} 
\langle \delta r^2 \rangle = \frac{1}{ N_f} \sum_{i=1}^{N_f} \int ds_i ~\frac{ \delta r_i^2(s_i)}{\ell_i} .
\end{equation}
where $N_f = 17$ is the number of filaments in the bundle and $\ell_i$ is the arc length used in the averaging of the $i$th filament.  We note that both quantities are insensitive to variations in spacing from pair to pair throughout the cross section (i.e. whether a packing is isometric or not), and only measure longitudinal variations.

We analyze filament bundles from tangent fields that are constructed to twist around a planar, circular central curve of radius $\kappa_0^{-1}$, the major radius of the torus, with a minor radius $R$, which is defined by the outer filament in the bundle.  As detailed below, for a general non-equidistant family of tangent fields, the winding rate of filaments around the minor cycle of the torus is non-uniform.  We therefore impose an additional constraint that all curves in the cross section have the same average circulation rate around the minor cycle of the torus.  In terms of the dependence of the angular position $\phi$ of a given curve (parameterized by the arc position $s$ along the central curve), this takes the form of constant pitch
\begin{equation}
P = 2 \pi/\Omega \equiv \int_{0}^{2 \pi} d \phi \Big( \frac{\partial \phi}{\partial s} \Big)^{-1} .
\label{eq: pitch}
\end{equation}
Using this definition of $\Omega$, we compare the uniformity of spacing in each ansatz as a function of reduced curvature $\kappa_0 R$ and reduced twist $\Omega R$.  When computing length averages, we average over the pitch length, or a half-circumference of the central circle $L = \pi/\kappa_0$ when $L/P<1$.

\subsection{$S^3$ fibrations projected to Euclidean space}

While there are no equidistant filament textures in $\mathbb{R}^3$ which are both twisted and bent, the same is not true of more general curved spaces.  In particular, $S^3$, the unit sphere in $\mathbb{R}^4$, permits a family of twisted, equidistant curves called Clifford parallels \cite{sadoc_geometrical_1999}.  These uniformly double-twisted
curves, which generate the Hopf fibration, are equidistant in $S^3$, but when stereographically projected into $\mathbb{R}^3$ induce a twisted, toroidal structure of interlinking circles.   Stereographic projections of the Hopf fibration to $\mathbb{R}^3$ generate twisted toroidal bundles with a particular linking number,  or ratio of bend to twist, $|\Omega|/\kappa_0 =1$.  Projection of a more general class of fibrations, the Seifert fibrations, which are also equidistant in $S^3$, permit a variable ratio of bend to twist~\cite{sadoc_3-sphere_2009, grason_colloquium_2015} .  Because stereographic projection preserves metric properties at the pole of the projection, which is chosen to be the major cycle at the center of the bundle, these projections of Seifert fibrations of $S^3$ have been proposed as physical models of cyclized, chiral polymer condensates that compromise between uniform packing and twist, \cite{kleman_frustration_1985, charvolin_geometrical_2008, mosseri_hopf_2012}.

Here, we construct projections of Seifert fibrations following the toroidal coordinates of Sadoc and Charvolin~\cite{sadoc_3-sphere_2009}.  With coordinates for the sphere in $\mathbb{R}^4$ of radius $ \kappa_0^{-1} $ given by
\begin{align}
x_1 &= \kappa_0^{-1} \cos{\varphi}\sin{\Theta} \nonumber \\
x_2 &=  \kappa_0^{-1}  \sin{\varphi}\sin{\Theta} \nonumber \\
x_3 &=  \kappa_0^{-1}  \cos{\psi}\cos{\Theta}  \\
x_4 &=  \kappa_0^{-1}  \sin{\psi}\cos{\Theta} \nonumber,
\end{align}
the fibers of a Seifert fibration are defined by $\varphi(\psi) = \varphi_0 + \alpha \psi$, where $\psi$ is a parameter that travels along the fibers and $\alpha$ parameterizes the ratio of turns per minor cycle of the torus to the turns per major cycle~\footnote{Strictly, $\alpha$ is a rational number such that the $(a,b)$ Seifert fibration of $S^{3}$ has $\alpha = \tfrac{a}{b}$.}.  The coordinate $\Theta$ parameterizes different tori, each of which is foliated by curves of distinct values of $\varphi_0 \in[0,2\pi]$.  Stereographically projecting a fiber to $\mathbb{R}^3$ through a pole of $S^3$ (where $\Theta=0$ corresponds to the major cycle of radius $\kappa_0^{-1}$ in $\mathbb{R}^3$) a fiber at $\Theta$ and $\phi_0$ parameterized by $\psi$ is given in Cartesian coordinates by
\begin{align}
x(\psi) &= \kappa_0^{-1} \frac{ \cos{\psi}\cos{\Theta}}{1-\cos{(\varphi_0 + \alpha \psi)}\sin{\Theta}} \nonumber \\
y(\psi) &= \kappa_0^{-1} \frac{ \sin{\psi}\cos{\Theta}}{1-\cos{(\varphi_0 + \alpha \psi)}\sin{\Theta}} \nonumber \\
z(\psi) &= \kappa_0^{-1}  \frac{ \sin{(\varphi_0 + \alpha \psi)}\sin{\Theta}}{1-\cos{(\varphi_0 + \alpha \psi)}\sin{\Theta}}.
\label{eq:seifertdef}
\end{align}
This projection is composed of curves defined on nested tori of increasing minor radius, $\kappa_0^{-1} \tan \Theta$.  However, the tori are not concentrically nested around a fixed major circle, and instead, are centered around major circles of increasing radius $\kappa_0^{-1} \sec \Theta$.  Because the arc distance along the central curve is simply $(\Delta \psi) \kappa_0^{-1}$, it is straightforward to see that the twist, as defined in Eq.~\eqref{eq: pitch}, is $\Omega = \alpha \kappa_0$.

Due to the non-concentric nature of toroidal stacking in this projection, it is convenient to analyze the tangent field in terms of orthonormal directions $\hat{ t}_0 =\kappa_0 \partial_\psi \mathbf{x}\big|_{\Theta =0}$,  $\hat{\varphi} = \partial_\varphi \mathbf{x}/|\partial_\varphi \mathbf{x}|$  and $\hat{\Theta} = \hat{t}_0 \times \hat{\varphi} $.  The tangent vector field of the texture induced by the Seifert fibers can now be found by differentiating Eq.~\eqref{eq:seifertdef}, with respect to $\psi$:
\begin{equation}
\mathbf{t} = \frac{ \cos{\Theta} ~\hat{t}_0  +  \alpha \sin{\Theta}~ \hat{\varphi}}{\sqrt{\cos^2{\Theta} + \alpha^2 \sin^2{\Theta}}}.
\end{equation}
From this, the components of $H_{\alpha \beta}$ along $\hat{\Theta}$ and $\hat{b} = {\mathbf t} \times \hat{\Theta}$ can be found explicitly:
\begin{align}
H_{\Theta \Theta} =H_{b b}  &=  - 2 \sqrt{2}   \Omega \frac{ \sin{ \Theta}\sin{\varphi }}{\sqrt{\cos^2{\Theta} + \alpha^2 \sin^2{\Theta}}} \nonumber \\
H_{\Theta b} &= H_{b \Theta} = 0 \nonumber .
\label{seifert_scale}
\end{align}
This diagonal structure of the convective flow of separation follows from the stereographic projection: relative to the equidistant fibrations in $S^3$, the local distances between curves is locally stretched by the projection to $\mathbb{R}^3$ by equal amounts in both directions normal to $\mathbf{t}$.  Qualitatively, the spatial variation of non-equidistance follows that illustrated for the $det({\bf H})=0$ structure in the Supplemental Video, with respective bunching and of filaments on the inner and outer sides of the torus.  While similar topology and spatial distribution of non-equidistance, we find that the magnitude of spacing variation differs considerably among the ansatzes.

We note that in the limit of narrow bundles ($\Theta \to 0$), we can estimate the growth of non-equidistance from $H \sim \Omega \kappa_0 \rho$.  When $\Omega \gg \kappa_0$ we average this over one $P$ (a minor cycle of the torus) to estimate $\delta r \sim \kappa_0 \rho$.  Alternatively, for small twist when $\Omega \ll \kappa_0$ this should be averaged over the bundle length $2 \pi/\kappa_0$, leading to $\delta r \sim \Omega \rho$.  From these two regimes, we estimate the scaling of non-equidistance with bundle thickness
\begin{equation}
\label{eq: dr_seifert}
\lim_{R \to 0} \langle \delta r^2 \rangle_{\text{Seifert}} \propto {\rm min}[\Omega^2,\kappa_0^2]\times R^2 .
\end{equation}
We compare this estimate to numerical calculations of $\langle \delta r^2 \rangle$ in the $\kappa_0 R$ and $\Omega R$ plane for projections of Seifert fibrations in Fig.~\ref{fig:seifert_numerics}.  

\subsection{Splay-free toroids}
The non-equidistance of stereographic projections of fibrations of $S^3$ derives from the locally isotropic (conformal) dilation of inter-filament spacing.  An alternative ansatz, and one which is typically invoked in models of polymeric liquid crystal textures, is the assumption of zero splay, which corresponds to constant area per filament transverse to its normal \cite{de_gennes_polymeric_1976}.  Hence, in the plane transverse to each filament, the polygonal region bounded by the neighboring filaments maintains constant area, and exhibits only area-preserving (shear) deformations as it flows along its contour.  

A splay-free tangent field requires that $\grad \cdot {\bf t}  =\text{tr}(\mathbf{H})$ vanishes.  Since $H_{\rho \rho} =0$ by construction in the generalized cylindrical coordinates of Sec.~\ref{equisol}, this imposes the additional condition that $H_{b b} =0$, or Eq.~\eqref{termsandconditions2}.  For a circular central curve, which has constant curvature and zero torsion, this equation can be solved by the method of characteristics,
giving:
\begin{equation}
\sin{\theta} = \frac{f(\rho)}{1-\rho \kappa_0 \cos{\phi}}, 
\label{kulic}
\end{equation}
where $f(\rho)$ is any function of $\rho$.  Previous studies for splay-free liquid crystalline toroids have assumed the simple linear ansatz, e.g. $f(\rho) = \Omega \rho$.  Notably, a splay-free toroidal texture is spatially limited to $f(\rho) + \rho \kappa_0 < 1$, beyond which it becomes singular.  The additional constraint that all curves wind around the minor cycle of the toroid at the same pitch, Eq.~\eqref{eq: pitch}, constrains the specific radial dependence of $f(\rho)$ and $\kappa_0 \rho$.  The rate of angular circulation of a filament's position relative to the inward pointing normal of the major circle is
\begin{equation}
\frac{ \partial \phi}{\partial s}  = \frac{f(\rho)}{\rho \sqrt{1-\frac{f(\rho)^2}{(1-\rho \kappa_0 \cos{\phi})^2}}}.
\label{eq:splayfreephidot}
\end{equation}
Inserting this into Eq.~\eqref{eq: pitch}, we have the additional condition that
\begin{equation}
\kappa_0 P = \int_{-\pi}^{+\pi} d \phi ~ \kappa_0 \rho \sqrt{\frac{1}{f(\rho)^2} - \frac{1}{(1-\rho \kappa_0 \cos{\phi})^2} } 
\label{eq:trfree_pitch}
\end{equation}
is independent of $\rho$.  From this condition, we derive the relationship between $f(\rho)$ and $\kappa_0 \rho$ for general values of $\kappa_0 P$ in splay-free bundles with mean winding which is notably more complex than the linear ansatz assumed in refs. \cite{kulic_twist-bend_2004,koning_saddle-splay_2014}.  Notably, in the slender bundle limit (as $ \Omega \rho\text{, } \kappa_0 \rho \to 0$). Eq.~\eqref{eq:trfree_pitch} satisfies
\begin{equation}
\label{eq: frho}
f(\rho) \simeq \frac{\Omega \rho}{\sqrt{1+ \Omega^2 \rho^2}}\big [1 - 18 \Omega^2 \kappa_0^2 \rho^4 + \mathcal{O}(\rho^6) \big] ,
\end{equation}
which, as $\kappa_0 \rho \to 0$, recovers the equidistant helical domains, for which $f(\rho) = \Omega \rho/\sqrt{1 + \Omega^2 \rho^2}$.

We can estimate the magnitude of this variable spacing by considering the off-diagonal, non-vanishing component of $H$ for Eq.~\eqref{kulic}, 
\begin{align}
H_{b \rho} = H_{\rho b} &= \partial_\rho \theta - \frac{\sin{\theta}\cos{\theta}}{\rho (1-\rho \kappa_0 \cos{\phi})} \nonumber \\
&=  \tan \theta \Big(\frac{\partial_\rho f}{f} +\frac{\kappa_0 \cos \phi}{f} \sin \theta\Big) - \frac{\sin{\theta}\cos{\theta}}{\rho (1-\rho \kappa_0 \cos{\phi})}  .
\label{kulic_scale}
\end{align}
In the limit of narrow splay-free bundles, we have  $H_{b \rho}\approx \Omega^3 \kappa_0 \rho^3 + {\cal O}(\rho^4)$.  Integrating this over the shorter of lengths $P$ and $L=\pi \kappa_0^{-1}$, we find that this separation averages to $\delta r \sim {\rm min}[\Omega, \kappa_0] \Omega^2 \rho^3$, from which we estimate,
\begin{equation}
\label{eq: dr_splayfree}
\lim_{R \to 0} \langle \delta r^2 \rangle_{splay-free}\propto {\rm min}[\Omega^2, \kappa_0^2]\times \Omega^4 R^6  .
\end{equation}
The suppression of splay notwithstanding, we find that the growth of spacing variation (shears) in narrow splay-free bundles grows as $\rho^3$, as opposed to the linear scaling with thickness of the stereographically projected fibrations of $S^3$~\footnote{It can be shown that neglect of the constant circulation constraint of Eq.~\eqref{eq:trfree_pitch} in the linear ansatz $f(\rho) = \Omega \rho$ studied in refs.~\cite{kulic_twist-bend_2004,koning_saddle-splay_2014} leads to less equidistanct splay-free textures, with $\delta r \sim {\rm min}[\Omega, \kappa_0] \Omega \rho^2$}.  Fig.~\ref{fig:trfree_numerics} shows the numerical calculation  of $\langle \delta r^2 \rangle$ in the $\kappa_0 R$ and $\Omega R$ plane for splay-free bundles.  Notably, due to the condition $f(R) \leq \kappa_0 R$, the continuous class of solutions extend only up to a critical thickness $R_{max} < \kappa_0$, whose value decreases with $\Omega R$.

\subsection{$\text{det}(\mathbf{H}) = 0$ toroids}
\begin{figure*}[h!]
\centering
\begin{subfigure}[t]{0.42\textwidth}
\includegraphics[height=7.25cm]{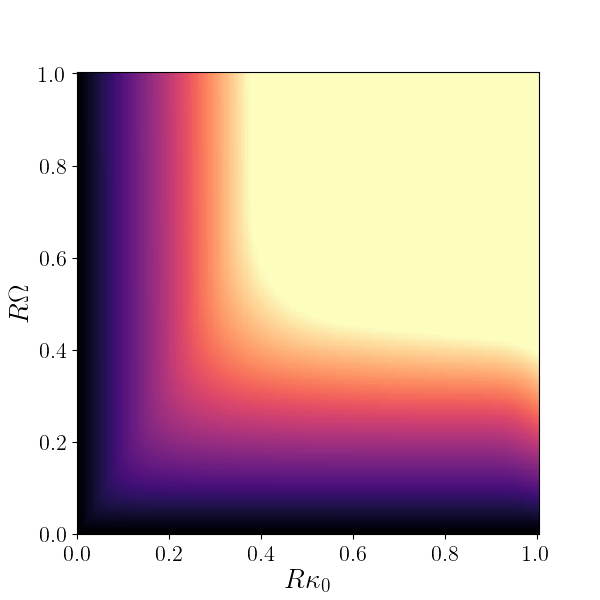}
\subcaption{}
\label{fig:seifert_numerics}
\end{subfigure}
\begin{subfigure}[t]{0.42\textwidth}
\includegraphics[height = 7.25cm]{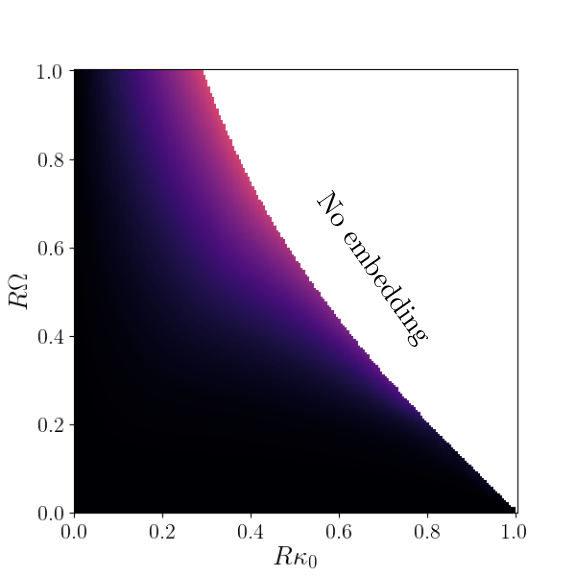}
\subcaption{}
\label{fig:trfree_numerics}
\end{subfigure}
\begin{subfigure}[t]{.1\textwidth}
\raisebox{.65cm}{\includegraphics[height=6cm]{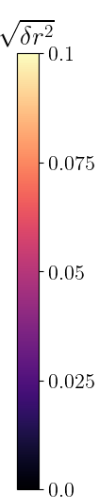}}
\end{subfigure}
\begin{subfigure}[t]{0.42\textwidth}
\raisebox{.45cm}{\includegraphics[height=7.25cm]{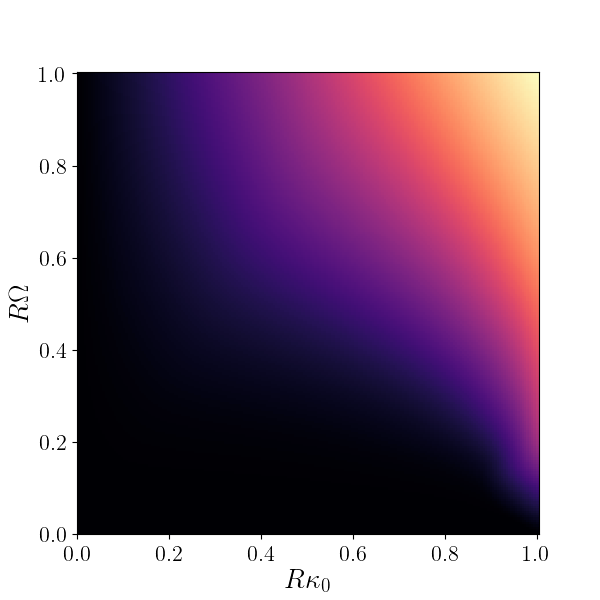}}
\subcaption{}
\label{fig:detfree_numerics}
\end{subfigure}
\begin{subfigure}[t]{0.42\textwidth}
\includegraphics[height=7.25cm]{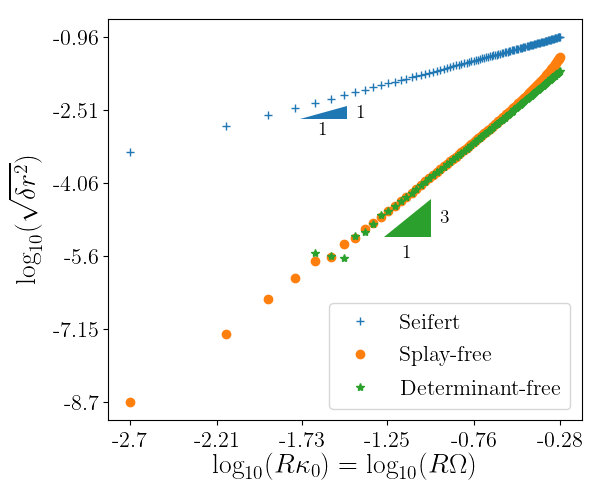}
\subcaption{}
\label{fig:scaling}
\end{subfigure}
\begin{subfigure}[t]{.1\textwidth}
\hskip .1\textwidth
\end{subfigure}
\caption{Numerically calculated deviations from equidistance for the Seifert fibrations~\ref{fig:seifert_numerics}, splay-free ($\text{Tr}(H) = 0$)~\ref{fig:trfree_numerics}, and $\text{det}(H) = 0$~\ref{fig:detfree_numerics} textures, varying $R \kappa_0$ and $R \Omega$, where $R$ is the bundle radius.
$R \Omega = R\kappa_0$ slices for Seifert, splay-free, and determinant-free structures, in \ref{fig:scaling} show, respectively, the $R$ scaling of the Seifert fibrations (Eq.~\eqref{eq: dr_seifert}) and the $R^3$ scaling of the splay-free (Eq.~\eqref{eq: dr_splayfree}), and determinant-free (Eq.~\eqref{eq: dr_detzero}) textures.}
\label{equidistantnumerics}
\end{figure*}

Finally, we consider a nearly-equidistant ansatz that satisfies $\text{det}(\mathbf{H}) = 0$, as opposed to vanishing trace.
In particular, we adopt the solution to $H_{\rho \rho} = H_{\rho b} =  H_{ b \rho} = 0$ of Eq.~\eqref{genericrhosolution}, and further take $\Omega$ to be a constant, such that the tangent field (in the coordinates of Eq.~\eqref{coorddef}) is
\begin{equation}
\tan \theta = \frac{\Omega \rho}{1-\kappa_0 \cos \phi} ,
\end{equation} 
which can be extended continuously up to thicknesses equal to the major radius of the torus.  Using the fact that $\tau_0$ and $\kappa_0$ are also constant, we find from Eq.~\eqref{eq:Hbb} that the non-vanishing component of ${\bf H}$ is 
\begin{equation}
H_{bb} =-\frac{2\rho^3 \kappa_0 \Omega^3 \sin{\phi}}{[(1-\rho \kappa_0 \cos{\phi})^2 + (\rho \Omega)^2]^{3/2}},
\label{atkinson_scale}
\end{equation}
which notably grows as $\sim \rho^3$ for small thicknesses.  Integrating over the shorter of $P$ or $\kappa_0$, we estimate the growth of non-equisdistance for this class of toroids to be
\begin{equation}
\label{eq: dr_detzero}
\lim_{R \to 0} \langle \delta r^2 \rangle_{{\rm det}({\bf H})=0}\propto {\rm min}[\Omega^2 ,\kappa_0^2 ]\times \Omega^4 R^6  .
\end{equation}
Thus, like the splay-free toroids, the ${\rm det}({\bf H}) =0$ ansatz remains more equidistant than $S^3$ fibrations (i.e. $\langle \delta r^2 \rangle^{1/2} \sim R^3$ as opposed to $\sim R$).  In Fig.~\ref{fig:scaling} we compare the numerical calculations for $\langle \delta r^2 \rangle^{1/2}$ for the three {\it ansatz} with $\Omega = \kappa_0$ for increasing twist.  For increasing thickness $\Omega R \lesssim 1$, we see that $\langle \delta r^2 \rangle$ ultimately grows larger for splay-free structures than for the  ${\rm det}({\bf H}) =0$ ansatz, indicating that the incorporation of a small amount of splay leads to more equidistant structures.  How close the ${\rm det}({\bf H}) =0$ structure comes to the true minimizer of $\langle \delta r^2 \rangle$ remains an open question.

\section{Discussion}
In this paper, we have presented several results on packings of multiple curves in $\mathbb{R}^3$ constrained by mutual equidistance.
First, we showed that any two mutually equidistant curves $\mathbf{r}_1$ and $\mathbf{r}_2$ in $\mathbb{R}^3$ are spanned by the ruled surface generated by the vector distance of closest approach between the curves, and the one parameter family of curves perpendicular to these rulings is itself equidistant.  We call this the {\it separating surface} defined by the equidistant pair, and between two equidistant curves, it is possible to fill in an arbitrary number of mutually equidistant curves embedded in the separating surface.  Although such families of curves are clearly unlimited in number, they are strictly two-dimensional in the sense that the family is collinear:  the 1D line separating any two curves perpendicularly intersects all the curves in the set.

In contrast, we find that non-collinear, volume filling, curve fields of $\mathbb{R}^3$ fall into two strictly distinct families, and in comparison to the collinear families, the geometries of curves that these permit are highly constrained.  Crudely speaking, bundles of curves can be twisted (uniformly) but not bent, or bent but not twisted.  However, like the collinear family, these equidistant curve fields have the property that they allow for embedding an arbitrary number of equidistant curves ($N \to \infty$) within a finite tubular neighborhood of some central curve in $\mathbb{R}^3$.  

The relatively restrictive geometry of equidistant fields raises interesting questions about the relationship between the problem of packing finite vs. infinite equidistant curves.  The existence of only two distinct equidistant fields, along with the tube argument in Fig. \ref{fig:tubeconstruction} suggests that the structure of finite $N$ equidistant bundles may be much less constrained than equidistant fields. Discrete equidistant bundles of this sort have ready applications to physical systems, from collagen triple helices
\cite{brodsky_molecular_2005} and other dense packed biological systems, to the (conventional) seven strands that make up most wire rope \cite{costello_theory_1990}.  A particularly relevant restriction of this problem is that of \emph{locally isometric} packings, where each filament is only constrained to lie equidistant to its nearest neighbors at some characteristic distance $a$, as in typical physical systems, where filaments packings are governed by an interfilament spacing set by an effective size.

We also expect the constraint satisfaction problem for $N$ equidistant filaments to yield novel and complex geometries,
since twisted equidistant triplets can be constructed around any smooth curve in $\mathbb{R}^3$, but only constant twist
helical bundles have a continuous field realization.  We conjecture that there exists $N_c>3$ such that the only bundles of $N\geq N_c$ regular,
equidistant, non-collinear curves in $\mathbb{R}^3$ are either parallel (developable domain) or helical (constant twist), i.e. they are integral curves of equidistant fields.

For relatively small numbers of filaments, ($N \leq 3$), these and related close packing problems have been studied in the context of ideal (or tight) knots
and tangles \cite{katritch_geometry_1996}.  Ideal knots, which are embeddings in $\mathbb{R}^3$ that minimize the ratio of knot length to filament-width \cite{calvo_biarcs_2005},
demand a fully global treatment that considers self-contact phenomena.  To this end, the principle object of study for single stranded
knots becomes not the distance of closest approach, but the global radius of curvature \cite{gonzalez_global_1999}.
Interestingly, ideal knot embeddings are not equidistant in general, even when equidistant embeddings exist.  For example, the ideal trefoil is known to make close (self-)contact over only a subset of its length \cite{ashton_knot_2011}.  The existence of geometrically rigid families of equidistant curve packings suggest that knot optimization problems that account for the energetic penalty of broken cohesive contacts are likely to yield new classes of minimizers \cite{calvo_physical_2005}.  For example, one may consider a generalization of the ``M\"obius energy'' \cite{janse_van_rensburg_tutorial_2005}, that incorporates a pair-wise potential between different arc-elements of a knotted curve, parameterized by some $V(x)$ that diverges as $x \to 0$,
\begin{equation}
E[\boldsymbol{\gamma}] = \int ds \int d\sigma V(|\boldsymbol{\gamma}(\sigma) - \boldsymbol{\gamma}(s)|) ,
\end{equation}
where  $s$ and $\sigma$ are arc length parameterizations of curve $\boldsymbol{\gamma}$.  When $V(x)$ is a strictly hard-core repulsive potential, we recover the ideal knot problem, while if $V(x)$ has an attractive minimum at finite $x= \delta$, we might expect solutions which favor equidistance.  In particular, in the limit that the cohesive attraction becomes infinitely strong in depth but infinitely narrow in range (relative to the repulsive core thickness), we anticipate a new class of minimizers that maximize the length and number of cohesive contacts.   In light of the conjectured rigidification of the constraints on equidistance  with increasing numbers of curves (or, here, curve segments) in equidistant contact, we further anticipate that such minimizers will be strongly dependent on the knot topology.  For example, because torus knots are necessarily simultaneously bent and twisted, we expect uniformly equidistant cohesive contact to be possible only when the number of strands arrayed around the minor cycle of the knot is less than $N_c$.

Beyond possible applications to problems in knot theory, the geometric constraints of equidistance would seem to have important and heretofore unexplored mechanical and structural consequences for a range of multi-filament structures.  Recent experimental studies, for example, have shown that 2D packings of initially straight filaments tend to adopt constant-pitch, helical shapes when subjected to mechanical twist at their ends \cite{panaitescu_measuring_2017}.  The emergence of this texture, even in the absence of cohesion between filaments, suggests that equidistance may be favored due to generic mechanical arguments (e.g. due to inward pressures generated by flexed or stretched outer strands).  This observation, in combination with the restrictive constraints imposed by equidistance in large $N$ packings, as described herein, raises further questions about the additional mechanical responses of filament packings associated with driving the structure to a non-equidistant geometry, such as when one simultaneously bends and twists a packing.   Bent and twisted assemblies of filaments, twisted toroids, are observed in condensates of collagen~\cite{ cooper_precipitation_1969} and DNA~\cite{leforestier_structure_2009}, and physical models constructed to date have yet to account for necessary energetic costs of non-equidistance required by this geometry.  

Beyond even structures of physical filament, twisted toroidal structures appear as topological solitons in range of classical field theories, for example, the extended non-linear $\sigma$ model \cite{faddeev_stable_1997,battye_knots_1998}, which supports knotted solutions whose topology is closely connected to the Hopf fibration of $S^3$.  In these ``hopfion" structures, 1D preimages of constant order parameter orientation (corresponding to a point on $S^2$) correspond to ``virtual filaments" that are twisted into closed toroidal bundles.  Above, we showed that the simultaneously twisted and bent structure of hopfions is incompatible with equidistance between preimages.  Recent studies show that hopfions emerge in models with preferred chiral pitch, such as models of chiral liquid crystals \cite{ackerman_diversity_2017,ackerman_static_2017}, and chiral \cite{sutcliffe_hopfions_2018, liu_binding_2018}, or frustrated ferromagnets \cite{sutcliffe_skyrmion_2017}.  In such models, a preferred rotation rate corresponds to a favored constant local spacing between preimage ``filaments'' of the field configuration.  Hence, we expect that equidistant (but not necessarily isometric) textures of constant-preimage filaments are energetically favored.  Thus, at least in models with a preferred twist wavelength, the incompatibility between twist, bend and equidistance in curve fields in $\mathbb{R}^3$ represents an intrinsic, and previously unrecognized, source of frustration in the formation of hopfionic structures.

Addressing questions about the structural and mechanical consequences for complex, non-equidistant bundle geometries requires new theoretical descriptions, since canonical approaches, such as the generalized elasticity theory of columnar liquid crystals \cite{degennes_physics_1995}, account for only small deviations around an unstrained reference.  The relevant physics for twisted and bent filament bundles (e.g. twisted toroids) requires a fully geometrically non-linear theory that couples the metric
properties of the cross-sectional filament packing to the flow generated by the filament texture, a framework which will be addressed in future work.

Of particular interest is the coupling of  metric (2D solid) to textural (1D fluid) degrees of freedom in geometrically frustrated materials.  In the simplest case of helical filament bundles, the increase in twist leads to an effective positively curved metric and the stability of excess 5-fold disclinations in an otherwise hexagonally-coordinated bundle \cite{bruss_non-euclidean_2012}.  The total integrated Gaussian curvature of a straight twisted bundle is $2 \pi$, implying a maximum number of six excess 5-fold defects  \cite{bruss_non-euclidean_2012}.  For combined twisted and bent geometries, such as a twisted toroid, a naive analysis of the ``local metric" induced in a planar cut of the bundle suggests that the effective integrated curvature of the section exceeds the value for the straight bundle, presumably implying that simultaneously twisting and bending a bundle increases the total number of defects in the ground state order.  It remains to be understood whether, and to what extent, this ``local" perspective on the metric structure in a give planar cut of a non-equidistant bundle truly underlies even a heuristic understanding of the coupling between defects and the 3D geometry of bundles beyond the equidistant cases studied so far.  

For straight filament bundles, similar work has shown that the introduction of packing defects can generate highly non-trivial textures in cohesive filament bundles, through their ability to reshape the ``target metric" of a filament packing from planar to non-Euclidean~\cite{bruss_defect-driven_2018}. This effect neatly demonstrates one important repercussion of our result in Section~\ref{sec:fields}:  that the response of positive and negative topological defects (5- and 7-fold disclinations in hexagonal packings) is highly asymmetric \emph{because} there is an equidistant field with positive effective curvature, while there are no equidistant fields with negative effective curvature.  The consequences of the restrictive nature of equidistance in bundles with negative curvature are therefore even more severe, as evidenced by the non-trivial elastic instabilities observed in simulated bundles with trapped negative disclinations.  A theoretical approach to predict equilibrium configurations of bundles whose target metrics (controlled by either distributions of defects or by patterns of inhomogeneous filament diameter) are incompatibile with equidistance remains an open challenge.

\section{Acknowledgements}
We are grateful to R. Kusner and participants of the ``Packing of Continua" workshop at the Aspen Center for Physics (NSF PHY 1607611) for numerous helpful discussions.  We also thank D. Hall and H. Wu for comments on this manuscript.  This work was supported by the National Science Foundation under grant Nos. DMR-1608862 and DMR-1507377.

\appendix
\section{Existence of Equidistant Pairs}
\label{appdx:existence}
Let $\mathbf{r}_1$ be a curve embedded in $\mathbb{R}^3$ with Darboux frame $\{\hat{T}, \hat{e}_1, \hat{e}_2 \}$, arc length $s$, and frame curvatures and torsion $\kappa_1$, $\kappa_2$, and $\tau_g$. Then any curve $\mathbf{r}_2$ parameterized by
\begin{equation}
\mathbf{r}_2(s) = \mathbf{r}_1(s) + \rho \bigg\{\cos{\big[\phi(s)\big]} \hat{e}_1(s) + \sin{\big[\phi(s)\big]} \hat{e}_2(s)\bigg\},
\end{equation}
is equidistant to $\mathbf{r}_1$.  To see why, note that for any such $\mathbf{r}_2$, the point of closest approach to $\mathbf{r}_2(s)$  on $\mathbf{r}_1$ is the corresponding point $\mathbf{r}_1(s)$. Then $\mathbf{r}_1$ and $\mathbf{r}_2$ are equidistant when $\partial_s \mathbf{r}_2 \cdot (\mathbf{r}_2 - \mathbf{r}_1) = 0$, as in Eq. \eqref{equidistance_eqn}.  Since 
\begin{equation}
\mathbf{r}_2 - \mathbf{r}_1 = \rho \bigg\{\cos{\big[\phi(s)\big]} \hat{e}_1(s) + \sin{\big[\phi(s)\big]} \hat{e}_2(s)\bigg\},
\end{equation}
all that remains is to show that $\partial_s \mathbf{r}_2$ is perpendicular to $\cos{\big[\phi(s)\big]} \hat{e}_1(s) + \sin{\big[\phi(s)\big]} \hat{e}_2(s)$. Since
\begin{equation}
\partial_s \mathbf{r}_2 = (1 - \rho \kappa_1 \cos{\phi} - \rho \kappa_2 \sin{\phi})\hat{T} + \rho (\partial_s \phi + \tau_g)(- \sin{\phi} \hat{e}_1 + \cos{\phi} \hat{e}_2),
\end{equation}
the two vectors are always orthogonal, so we have that $\mathbf{r}_2$ and $\mathbf{r}_1$ are equidistant whenever
$(1 - \rho \kappa_1 \cos{\phi} - \rho \kappa_2 \sin{\phi})$ and $\rho (\partial_s \phi + \tau_g)$ are finite and nonzero.

\section{Quasi-cylindrical coordinates for filament bundles}
\label{appdx:coords}
We can write down any generic position $\vec{x}$ in coordinates centered around some curve $\mathbf{r}_0$ as follows:
\begin{equation}
\mathbf{x} = \mathbf{r}_0 + \rho \hat{\rho}.
\end{equation}
An infinitesimal displacement $d\vec{x}$ can then be found by
\begin{equation}
d\mathbf{x} = \frac{\partial \mathbf{x}}{\partial s}ds + \frac{\partial \mathbf{x}}{\partial \rho} d\rho + \frac{\partial \mathbf{x}}{\partial \phi} d\phi,
\end{equation}
where the partial derivatives are:
\begin{align}
\frac{\partial \mathbf{x}}{\partial s} &= \hat{t}_0 + \rho \frac{\partial{\hat{\rho}}}{{\partial s}} \\
\frac{\partial \mathbf{x}}{\partial \rho} &= \hat{\rho} \\
\frac{\partial \mathbf{x}}{\partial \phi} &= \rho \hat{\phi},
\end{align}
and
\begin{equation}
\frac{\partial \hat{\rho}}{\partial s} = - \kappa_0\cos{\phi}\hat{t}_0 + \tau_0 \hat{\phi}.
\end{equation}

So, we find the Jacobian for this coordinate transformation:
\begin{equation}
J =
\begin{bmatrix} 1 - \rho\kappa_0 \cos{\phi} & 0 & 0 \\
0 & 1 & 0 \\
\rho \tau_0 & 0 & \rho
\end{bmatrix}
\end{equation}
with its inverse
\begin{equation}
J^{-1} = \begin{bmatrix} \frac{1}{1-\rho\kappa_0 \cos{\phi}} & 0 & 0 \\
0 & 1 & 0 \\
-\frac{\tau_0}{1 - \rho \kappa_0 \cos{\phi}} & 0 & \frac{1}{\rho}.
\end{bmatrix}
\end{equation}
Note that this inverse does not exist for $\rho \kappa_0 \cos{\phi} = 1$ or $\rho = 0$, for which we can't take these derivatives.

We can now write down the tensor $\nabla \mathbf{t}$ in these coordinates, represented in the basis
$\{\mathbf{t}, \hat{\rho}, \hat{b}\}$, where $\hat{b} = \mathbf{t} \times \hat{\rho}$.
The $\rho \rho$ component of this matrix,
$\hat{\rho} \cdot \grad = \frac{\partial}{\partial \rho}$ doesn't do much, but  $\hat{e} \cdot \grad$ is slightly more exciting:
\begin{align}
\hat{e} \cdot \grad &= (J^{-1}\hat{e})\cdot \grad \\
&= \Bigg( \begin{bmatrix} \frac{1}{1-\rho\kappa_0 \cos{\phi}} & 0 & 0 \\
0 & 1 & 0 \\
-\frac{\tau_0}{1 - \rho \kappa_0 \cos{\phi}} & 0 & \frac{1}{\rho}
\end{bmatrix}
\cdot
\begin{bmatrix} -\sin{\theta} \\
0\\
\cos{\theta} \end{bmatrix}\Bigg) \cdot \grad \\
&= [- \sin{\theta} \frac{1}{1- \rho \kappa_0 \cos{\phi}} \hat{t}_0 + (\sin{\theta}\frac{\tau_0}{1-\rho \kappa_0 \cos{\phi}}
+ \cos{\theta} \frac{1}{\rho}) \hat{\phi}] \cdot \grad \\
&= - \sin{\theta} \frac{1}{1- \rho \kappa_0 \cos{\phi}} \frac{\partial}{\partial s}
+ (\sin{\theta}\frac{\tau_0}{1-\rho \kappa_0 \cos{\phi}} + \cos{\theta} \frac{1}{\rho}) \frac{\partial}{\partial \phi}.
\end{align}

We can now find these derivatives acting on the tangent field, noting that to find derivatives on $\hat{\phi}$,
we can write it explicitly in the Frenet-Serret frame $\hat{\phi} = -\sin{\phi} \hat{N}_0 + \cos{\phi} \hat{B}_0$, with
\begin{align}
\partial_s\hat{N}_0 &= -\kappa_{0} \hat{t}_0 + \tau_0 \hat{B}_0 \\
\partial_s{B}_0 &= -\tau_0 \hat{N}_0 \\
\implies \partial_s \hat{\phi} &= \kappa_0\sin{\phi} \hat{t}_0 - \tau_0 \hat{\rho}.
\end{align}
This gives us derivatives as follows:
\begin{align}
\partial_\phi \hat{t} &= \partial_\phi (\cos{\theta}\hat{t}_0) + \partial_\phi (\sin{\theta} \hat{\phi} \\
&= - \sin{\theta}\partial_\phi \theta \hat{t}_0 + \cos{\theta} \partial_\phi \theta - \sin{\theta} \hat{\rho}
\end{align}
\begin{equation}
\partial_s \hat{t} = \cos{\theta} \kappa_0(\cos{\phi} \hat{\rho} - \sin{\phi} \hat{\phi}) + \sin{\theta}(\kappa_0 \hat{t}_0 - \tau_0 \hat{\rho})
- \sin{\theta} \partial_s \theta \hat{t}_0 + \cos{\theta} \partial_s \theta \hat{\phi}
\end{equation}
\begin{equation}
\partial_\rho \hat{t} = -\sin{\theta} \partial_\rho \theta \hat{t}_0 + \cos{\theta} \partial_\rho \theta \hat{\phi}
\end{equation}
These now let us write down explicitly the components of $\nabla \mathbf{t}$, and give us Eq.~\eqref{gradttensor}.

\bibliography{Atkinson_2019}

\begin{thebibliography}{75}%
\makeatletter
\providecommand \@ifxundefined [1]{%
 \@ifx{#1\undefined}
}%
\providecommand \@ifnum [1]{%
 \ifnum #1\expandafter \@firstoftwo
 \else \expandafter \@secondoftwo
 \fi
}%
\providecommand \@ifx [1]{%
 \ifx #1\expandafter \@firstoftwo
 \else \expandafter \@secondoftwo
 \fi
}%
\providecommand \natexlab [1]{#1}%
\providecommand \enquote  [1]{``#1''}%
\providecommand \bibnamefont  [1]{#1}%
\providecommand \bibfnamefont [1]{#1}%
\providecommand \citenamefont [1]{#1}%
\providecommand \href@noop [0]{\@secondoftwo}%
\providecommand \href [0]{\begingroup \@sanitize@url \@href}%
\providecommand \@href[1]{\@@startlink{#1}\@@href}%
\providecommand \@@href[1]{\endgroup#1\@@endlink}%
\providecommand \@sanitize@url [0]{\catcode `\\12\catcode `\$12\catcode
  `\&12\catcode `\#12\catcode `\^12\catcode `\_12\catcode `\%12\relax}%
\providecommand \@@startlink[1]{}%
\providecommand \@@endlink[0]{}%
\providecommand \url  [0]{\begingroup\@sanitize@url \@url }%
\providecommand \@url [1]{\endgroup\@href {#1}{\urlprefix }}%
\providecommand \urlprefix  [0]{URL }%
\providecommand \Eprint [0]{\href }%
\providecommand \doibase [0]{http://dx.doi.org/}%
\providecommand \selectlanguage [0]{\@gobble}%
\providecommand \bibinfo  [0]{\@secondoftwo}%
\providecommand \bibfield  [0]{\@secondoftwo}%
\providecommand \translation [1]{[#1]}%
\providecommand \BibitemOpen [0]{}%
\providecommand \bibitemStop [0]{}%
\providecommand \bibitemNoStop [0]{.\EOS\space}%
\providecommand \EOS [0]{\spacefactor3000\relax}%
\providecommand \BibitemShut  [1]{\csname bibitem#1\endcsname}%
\let\auto@bib@innerbib\@empty
\bibitem [{\citenamefont {Aste}\ and\ \citenamefont
  {Weaire}(2008)}]{aste_pursuit_2008}%
  \BibitemOpen
  \bibfield  {author} {\bibinfo {author} {\bibfnamefont {T.}~\bibnamefont
  {Aste}}\ and\ \bibinfo {author} {\bibfnamefont {D.~L.}\ \bibnamefont
  {Weaire}},\ }\href {http://www.crcnetbase.com/isbn/9781420068184} {\emph
  {\bibinfo {title} {The Pursuit of Perfect Packing}}}\ (\bibinfo  {publisher}
  {Taylor \& Francis},\ \bibinfo {address} {New York},\ \bibinfo {year}
  {2008})\BibitemShut {NoStop}%
\bibitem [{\citenamefont {Conway}\ and\ \citenamefont
  {Sloane}(1998)}]{conway_sphere_1998}%
  \BibitemOpen
  \bibfield  {author} {\bibinfo {author} {\bibfnamefont {J.}~\bibnamefont
  {Conway}}\ and\ \bibinfo {author} {\bibfnamefont {N.~J.~A.}\ \bibnamefont
  {Sloane}},\ }\href@noop {} {\emph {\bibinfo {title} {Sphere {Packings},
  {Lattices} and {Groups}}}}\ (\bibinfo  {publisher} {Springer Science \&
  Business Media},\ \bibinfo {year} {1998})\BibitemShut {NoStop}%
\bibitem [{\citenamefont {Kl\'{e}man}(1983)}]{kleman_points_1983}%
  \BibitemOpen
  \bibfield  {author} {\bibinfo {author} {\bibfnamefont {M.}~\bibnamefont
  {Kl\'{e}man}},\ }\href@noop {} {\emph {\bibinfo {title} {Points, lines, and
  walls: in liquid crystals, magnetic systems, and various ordered media}}}\
  (\bibinfo  {publisher} {J. Wiley},\ \bibinfo {address} {Chichester; New
  York},\ \bibinfo {year} {1983})\BibitemShut {NoStop}%
\bibitem [{\citenamefont {Banavar}\ and\ \citenamefont
  {Maritan}(2003)}]{banavar_colloquium_2003}%
  \BibitemOpen
  \bibfield  {author} {\bibinfo {author} {\bibfnamefont {J.~R.}\ \bibnamefont
  {Banavar}}\ and\ \bibinfo {author} {\bibfnamefont {A.}~\bibnamefont
  {Maritan}},\ }\href {\doibase 10.1103/RevModPhys.75.23} {\bibfield  {journal}
  {\bibinfo  {journal} {Reviews of Modern Physics}\ }\textbf {\bibinfo {volume}
  {75}},\ \bibinfo {pages} {23} (\bibinfo {year} {2003})}\BibitemShut {NoStop}%
\bibitem [{\citenamefont {Banavar}\ \emph {et~al.}(2003)\citenamefont
  {Banavar}, \citenamefont {Gonzalez}, \citenamefont {Maddocks},\ and\
  \citenamefont {Maritan}}]{banavar_self-interactions_2003}%
  \BibitemOpen
  \bibfield  {author} {\bibinfo {author} {\bibfnamefont {J.~R.}\ \bibnamefont
  {Banavar}}, \bibinfo {author} {\bibfnamefont {O.}~\bibnamefont {Gonzalez}},
  \bibinfo {author} {\bibfnamefont {J.~H.}\ \bibnamefont {Maddocks}}, \ and\
  \bibinfo {author} {\bibfnamefont {A.}~\bibnamefont {Maritan}},\ }\href
  {\doibase 10.1023/A:1021010526495} {\bibfield  {journal} {\bibinfo  {journal}
  {Journal of Statistical Physics}\ }\textbf {\bibinfo {volume} {110}},\
  \bibinfo {pages} {35} (\bibinfo {year} {2003})}\BibitemShut {NoStop}%
\bibitem [{\citenamefont {Neukirch}\ and\ \citenamefont {van~der
  Heijden}(2002)}]{neukirch_geometry_2002}%
  \BibitemOpen
  \bibfield  {author} {\bibinfo {author} {\bibfnamefont {S.}~\bibnamefont
  {Neukirch}}\ and\ \bibinfo {author} {\bibfnamefont {G.}~\bibnamefont {van~der
  Heijden}},\ }\href {\doibase 10.1023/A:1027390700610} {\bibfield  {journal}
  {\bibinfo  {journal} {Journal of Elasticity}\ }\textbf {\bibinfo {volume}
  {69}},\ \bibinfo {pages} {41} (\bibinfo {year} {2002})}\BibitemShut {NoStop}%
\bibitem [{\citenamefont {Bozec}\ \emph {et~al.}(2007)\citenamefont {Bozec},
  \citenamefont {van~der Heijden},\ and\ \citenamefont
  {Horton}}]{bozec_collagen_2007}%
  \BibitemOpen
  \bibfield  {author} {\bibinfo {author} {\bibfnamefont {L.}~\bibnamefont
  {Bozec}}, \bibinfo {author} {\bibfnamefont {G.}~\bibnamefont {van~der
  Heijden}}, \ and\ \bibinfo {author} {\bibfnamefont {M.}~\bibnamefont
  {Horton}},\ }\href {\doibase 10.1529/biophysj.106.085704} {\bibfield
  {journal} {\bibinfo  {journal} {Biophysical Journal}\ }\textbf {\bibinfo
  {volume} {92}},\ \bibinfo {pages} {70} (\bibinfo {year} {2007})}\BibitemShut
  {NoStop}%
\bibitem [{\citenamefont {Bohr}\ and\ \citenamefont
  {Olsen}(2011)}]{bohr_close-packed_2011}%
  \BibitemOpen
  \bibfield  {author} {\bibinfo {author} {\bibfnamefont {J.}~\bibnamefont
  {Bohr}}\ and\ \bibinfo {author} {\bibfnamefont {K.}~\bibnamefont {Olsen}},\
  }\href {\doibase 10.1007/s00214-010-0761-3} {\bibfield  {journal} {\bibinfo
  {journal} {Theoretical Chemistry Accounts}\ }\textbf {\bibinfo {volume}
  {130}},\ \bibinfo {pages} {1095} (\bibinfo {year} {2011})}\BibitemShut
  {NoStop}%
\bibitem [{\citenamefont {Snir}\ and\ \citenamefont
  {Kamien}(2005)}]{snir_entropically_2005}%
  \BibitemOpen
  \bibfield  {author} {\bibinfo {author} {\bibfnamefont {Y.}~\bibnamefont
  {Snir}}\ and\ \bibinfo {author} {\bibfnamefont {R.~D.}\ \bibnamefont
  {Kamien}},\ }\href {\doibase 10.1126/science.1106243} {\bibfield  {journal}
  {\bibinfo  {journal} {Science}\ }\textbf {\bibinfo {volume} {307}},\ \bibinfo
  {pages} {1067} (\bibinfo {year} {2005})}\BibitemShut {NoStop}%
\bibitem [{\citenamefont {Bico}\ \emph {et~al.}(2004)\citenamefont {Bico},
  \citenamefont {Roman}, \citenamefont {Moulin},\ and\ \citenamefont
  {Boudaoud}}]{bico_adhesion_2004}%
  \BibitemOpen
  \bibfield  {author} {\bibinfo {author} {\bibfnamefont {J.}~\bibnamefont
  {Bico}}, \bibinfo {author} {\bibfnamefont {B.}~\bibnamefont {Roman}},
  \bibinfo {author} {\bibfnamefont {L.}~\bibnamefont {Moulin}}, \ and\ \bibinfo
  {author} {\bibfnamefont {A.}~\bibnamefont {Boudaoud}},\ }\href {\doibase
  10.1038/432690a} {\bibfield  {journal} {\bibinfo  {journal} {Nature}\
  }\textbf {\bibinfo {volume} {432}},\ \bibinfo {pages} {690} (\bibinfo {year}
  {2004})}\BibitemShut {NoStop}%
\bibitem [{\citenamefont {Zhang}\ \emph {et~al.}(2004)\citenamefont {Zhang},
  \citenamefont {Atkinson},\ and\ \citenamefont
  {Baughman}}]{zhang_multifunctional_2004}%
  \BibitemOpen
  \bibfield  {author} {\bibinfo {author} {\bibfnamefont {M.}~\bibnamefont
  {Zhang}}, \bibinfo {author} {\bibfnamefont {K.~R.}\ \bibnamefont {Atkinson}},
  \ and\ \bibinfo {author} {\bibfnamefont {R.~H.}\ \bibnamefont {Baughman}},\
  }\href {\doibase 10.1126/science.1104276} {\bibfield  {journal} {\bibinfo
  {journal} {Science}\ }\textbf {\bibinfo {volume} {306}},\ \bibinfo {pages}
  {1358} (\bibinfo {year} {2004})}\BibitemShut {NoStop}%
\bibitem [{\citenamefont {Thess}\ \emph {et~al.}(1996)\citenamefont {Thess},
  \citenamefont {Lee}, \citenamefont {Nikolaev}, \citenamefont {Dai},
  \citenamefont {Petit}, \citenamefont {Robert}, \citenamefont {Xu},
  \citenamefont {Lee}, \citenamefont {Kim}, \citenamefont {Rinzler},
  \citenamefont {Colbert}, \citenamefont {Scuseria}, \citenamefont {Tománek},
  \citenamefont {Fischer},\ and\ \citenamefont
  {Smalley}}]{thess_crystalline_1996}%
  \BibitemOpen
  \bibfield  {author} {\bibinfo {author} {\bibfnamefont {A.}~\bibnamefont
  {Thess}}, \bibinfo {author} {\bibfnamefont {R.}~\bibnamefont {Lee}}, \bibinfo
  {author} {\bibfnamefont {P.}~\bibnamefont {Nikolaev}}, \bibinfo {author}
  {\bibfnamefont {H.}~\bibnamefont {Dai}}, \bibinfo {author} {\bibfnamefont
  {P.}~\bibnamefont {Petit}}, \bibinfo {author} {\bibfnamefont
  {J.}~\bibnamefont {Robert}}, \bibinfo {author} {\bibfnamefont
  {C.}~\bibnamefont {Xu}}, \bibinfo {author} {\bibfnamefont {Y.~H.}\
  \bibnamefont {Lee}}, \bibinfo {author} {\bibfnamefont {S.~G.}\ \bibnamefont
  {Kim}}, \bibinfo {author} {\bibfnamefont {A.~G.}\ \bibnamefont {Rinzler}},
  \bibinfo {author} {\bibfnamefont {D.~T.}\ \bibnamefont {Colbert}}, \bibinfo
  {author} {\bibfnamefont {G.~E.}\ \bibnamefont {Scuseria}}, \bibinfo {author}
  {\bibfnamefont {D.}~\bibnamefont {Tománek}}, \bibinfo {author}
  {\bibfnamefont {J.~E.}\ \bibnamefont {Fischer}}, \ and\ \bibinfo {author}
  {\bibfnamefont {R.~E.}\ \bibnamefont {Smalley}},\ }\href {\doibase
  10.1126/science.273.5274.483} {\bibfield  {journal} {\bibinfo  {journal}
  {Science}\ }\textbf {\bibinfo {volume} {273}},\ \bibinfo {pages} {483}
  (\bibinfo {year} {1996})}\BibitemShut {NoStop}%
\bibitem [{\citenamefont {Livolant}\ \emph {et~al.}(1989)\citenamefont
  {Livolant}, \citenamefont {Levelut}, \citenamefont {Doucet},\ and\
  \citenamefont {Benoit}}]{livolant_highly_1989}%
  \BibitemOpen
  \bibfield  {author} {\bibinfo {author} {\bibfnamefont {F.}~\bibnamefont
  {Livolant}}, \bibinfo {author} {\bibfnamefont {A.~M.}\ \bibnamefont
  {Levelut}}, \bibinfo {author} {\bibfnamefont {J.}~\bibnamefont {Doucet}}, \
  and\ \bibinfo {author} {\bibfnamefont {J.~P.}\ \bibnamefont {Benoit}},\
  }\href {http://dx.doi.org/10.1038/339724a0} {\bibfield  {journal} {\bibinfo
  {journal} {Nature}\ }\textbf {\bibinfo {volume} {339}},\ \bibinfo {pages}
  {724} (\bibinfo {year} {1989})}\BibitemShut {NoStop}%
\bibitem [{\citenamefont {Costello}(1990)}]{costello_theory_1990}%
  \BibitemOpen
  \bibfield  {author} {\bibinfo {author} {\bibfnamefont {G.~A.}\ \bibnamefont
  {Costello}},\ }\href {\doibase 10.1007/978-1-4684-0350-3} {\emph {\bibinfo
  {title} {Theory of Wire Rope}}},\ Mechanical {Engineering} {Series}\
  (\bibinfo  {publisher} {Springer-Verlag, New York},\ \bibinfo {year}
  {1990})\BibitemShut {NoStop}%
\bibitem [{\citenamefont {Gilbert}(1979)}]{gilbert_packing_1979}%
  \BibitemOpen
  \bibfield  {author} {\bibinfo {author} {\bibfnamefont {E.~N.}\ \bibnamefont
  {Gilbert}},\ }\href {\doibase 10.1002/j.1538-7305.1979.tb02960.x} {\bibfield
  {journal} {\bibinfo  {journal} {The Bell System Technical Journal}\ }\textbf
  {\bibinfo {volume} {58}},\ \bibinfo {pages} {2143} (\bibinfo {year}
  {1979})}\BibitemShut {NoStop}%
\bibitem [{\citenamefont {Santangelo}\ \emph {et~al.}(2007)\citenamefont
  {Santangelo}, \citenamefont {Vitelli}, \citenamefont {Kamien},\ and\
  \citenamefont {Nelson}}]{santangelo_geometric_2007}%
  \BibitemOpen
  \bibfield  {author} {\bibinfo {author} {\bibfnamefont {C.~D.}\ \bibnamefont
  {Santangelo}}, \bibinfo {author} {\bibfnamefont {V.}~\bibnamefont {Vitelli}},
  \bibinfo {author} {\bibfnamefont {R.~D.}\ \bibnamefont {Kamien}}, \ and\
  \bibinfo {author} {\bibfnamefont {D.~R.}\ \bibnamefont {Nelson}},\ }\href
  {\doibase 10.1103/PhysRevLett.99.017801} {\bibfield  {journal} {\bibinfo
  {journal} {Physical Review Letters}\ }\textbf {\bibinfo {volume} {99}}
  (\bibinfo {year} {2007}),\ 10.1103/PhysRevLett.99.017801}\BibitemShut
  {NoStop}%
\bibitem [{\citenamefont {Kn{\"o}ppel}\ \emph {et~al.}(2015)\citenamefont
  {Kn{\"o}ppel}, \citenamefont {Crane}, \citenamefont {Pinkall},\ and\
  \citenamefont {Schr{\"o}der}}]{knoppel_stripe_2015}%
  \BibitemOpen
  \bibfield  {author} {\bibinfo {author} {\bibfnamefont {F.}~\bibnamefont
  {Kn{\"o}ppel}}, \bibinfo {author} {\bibfnamefont {K.}~\bibnamefont {Crane}},
  \bibinfo {author} {\bibfnamefont {U.}~\bibnamefont {Pinkall}}, \ and\
  \bibinfo {author} {\bibfnamefont {P.}~\bibnamefont {Schr{\"o}der}},\ }\href
  {\doibase 10.1145/2767000} {\bibfield  {journal} {\bibinfo  {journal} {ACM
  Trans. Graph.}\ }\textbf {\bibinfo {volume} {34}},\ \bibinfo {pages} {39:1}
  (\bibinfo {year} {2015})}\BibitemShut {NoStop}%
\bibitem [{\citenamefont {Kl{\'e}man}(1980)}]{kleman_developable_1980}%
  \BibitemOpen
  \bibfield  {author} {\bibinfo {author} {\bibfnamefont {M.}~\bibnamefont
  {Kl{\'e}man}},\ }\href {\doibase 10.1051/jphys:01980004107073700} {\bibfield
  {journal} {\bibinfo  {journal} {Journal de Physique}\ }\textbf {\bibinfo
  {volume} {41}},\ \bibinfo {pages} {737} (\bibinfo {year} {1980})}\BibitemShut
  {NoStop}%
\bibitem [{\citenamefont {Starostin}(2006)}]{starostin_perfect_2006}%
  \BibitemOpen
  \bibfield  {author} {\bibinfo {author} {\bibfnamefont {E.~L.}\ \bibnamefont
  {Starostin}},\ }\href {\doibase 10.1088/0953-8984/18/14/S04} {\bibfield
  {journal} {\bibinfo  {journal} {Journal of Physics: Condensed Matter}\
  }\textbf {\bibinfo {volume} {18}},\ \bibinfo {pages} {S187} (\bibinfo {year}
  {2006})}\BibitemShut {NoStop}%
\bibitem [{\citenamefont {Bruss}\ and\ \citenamefont
  {Grason}(2012)}]{bruss_non-euclidean_2012}%
  \BibitemOpen
  \bibfield  {author} {\bibinfo {author} {\bibfnamefont {I.~R.}\ \bibnamefont
  {Bruss}}\ and\ \bibinfo {author} {\bibfnamefont {G.~M.}\ \bibnamefont
  {Grason}},\ }\href@noop {} {\bibfield  {journal} {\bibinfo  {journal}
  {Proceedings of the National Academy of Sciences}\ }\textbf {\bibinfo
  {volume} {109}},\ \bibinfo {pages} {10781} (\bibinfo {year}
  {2012})}\BibitemShut {NoStop}%
\bibitem [{\citenamefont {Bouligand}(1980)}]{bouligand_geometry_1980}%
  \BibitemOpen
  \bibfield  {author} {\bibinfo {author} {\bibfnamefont {Y.}~\bibnamefont
  {Bouligand}},\ }\href {\doibase 10.1051/jphys:0198000410110129700} {\bibfield
   {journal} {\bibinfo  {journal} {Journal de Physique}\ }\textbf {\bibinfo
  {volume} {41}},\ \bibinfo {pages} {1297} (\bibinfo {year}
  {1980})}\BibitemShut {NoStop}%
\bibitem [{\citenamefont {Hud}\ and\ \citenamefont
  {Downing}(2001)}]{hud_cryoelectron_2001}%
  \BibitemOpen
  \bibfield  {author} {\bibinfo {author} {\bibfnamefont {N.~V.}\ \bibnamefont
  {Hud}}\ and\ \bibinfo {author} {\bibfnamefont {K.~H.}\ \bibnamefont
  {Downing}},\ }\href@noop {} {\bibfield  {journal} {\bibinfo  {journal}
  {Proceedings of the National Academy of Sciences}\ }\textbf {\bibinfo
  {volume} {98}},\ \bibinfo {pages} {14925} (\bibinfo {year}
  {2001})}\BibitemShut {NoStop}%
\bibitem [{\citenamefont {Leforestier}\ and\ \citenamefont
  {Livolant}(2009)}]{leforestier_structure_2009}%
  \BibitemOpen
  \bibfield  {author} {\bibinfo {author} {\bibfnamefont {A.}~\bibnamefont
  {Leforestier}}\ and\ \bibinfo {author} {\bibfnamefont {F.}~\bibnamefont
  {Livolant}},\ }\href {\doibase 10.1073/pnas.0901240106} {\bibfield  {journal}
  {\bibinfo  {journal} {Proceedings of the National Academy of Sciences}\
  }\textbf {\bibinfo {volume} {106}},\ \bibinfo {pages} {9157} (\bibinfo {year}
  {2009})}\BibitemShut {NoStop}%
\bibitem [{\citenamefont {Cooper}(1969)}]{cooper_precipitation_1969}%
  \BibitemOpen
  \bibfield  {author} {\bibinfo {author} {\bibfnamefont {A.}~\bibnamefont
  {Cooper}},\ }\href {https://www.ncbi.nlm.nih.gov/pmc/articles/PMC1187741/}
  {\bibfield  {journal} {\bibinfo  {journal} {Biochemical Journal}\ }\textbf
  {\bibinfo {volume} {112}},\ \bibinfo {pages} {515} (\bibinfo {year}
  {1969})}\BibitemShut {NoStop}%
\bibitem [{\citenamefont {Ackerman}\ and\ \citenamefont
  {Smalyukh}(2017{\natexlab{a}})}]{ackerman_diversity_2017}%
  \BibitemOpen
  \bibfield  {author} {\bibinfo {author} {\bibfnamefont {P.~J.}\ \bibnamefont
  {Ackerman}}\ and\ \bibinfo {author} {\bibfnamefont {I.~I.}\ \bibnamefont
  {Smalyukh}},\ }\href {\doibase 10.1103/PhysRevX.7.011006} {\bibfield
  {journal} {\bibinfo  {journal} {Physical Review X}\ }\textbf {\bibinfo
  {volume} {7}},\ \bibinfo {pages} {011006} (\bibinfo {year}
  {2017}{\natexlab{a}})}\BibitemShut {NoStop}%
\bibitem [{\citenamefont {Ackerman}\ and\ \citenamefont
  {Smalyukh}(2017{\natexlab{b}})}]{ackerman_static_2017}%
  \BibitemOpen
  \bibfield  {author} {\bibinfo {author} {\bibfnamefont {P.~J.}\ \bibnamefont
  {Ackerman}}\ and\ \bibinfo {author} {\bibfnamefont {I.~I.}\ \bibnamefont
  {Smalyukh}},\ }\href {\doibase 10.1038/nmat4826} {\bibfield  {journal}
  {\bibinfo  {journal} {Nature Materials}\ }\textbf {\bibinfo {volume} {16}},\
  \bibinfo {pages} {426} (\bibinfo {year} {2017}{\natexlab{b}})}\BibitemShut
  {NoStop}%
\bibitem [{\citenamefont {Sutcliffe}(2017)}]{sutcliffe_skyrmion_2017}%
  \BibitemOpen
  \bibfield  {author} {\bibinfo {author} {\bibfnamefont {P.}~\bibnamefont
  {Sutcliffe}},\ }\href {\doibase 10.1103/PhysRevLett.118.247203} {\bibfield
  {journal} {\bibinfo  {journal} {Physical Review Letters}\ }\textbf {\bibinfo
  {volume} {118}} (\bibinfo {year} {2017}),\
  10.1103/PhysRevLett.118.247203}\BibitemShut {NoStop}%
\bibitem [{\citenamefont {Sutcliffe}(2018)}]{sutcliffe_hopfions_2018}%
  \BibitemOpen
  \bibfield  {author} {\bibinfo {author} {\bibfnamefont {P.}~\bibnamefont
  {Sutcliffe}},\ }\href {\doibase 10.1088/1751-8121/aad521} {\bibfield
  {journal} {\bibinfo  {journal} {Journal of Physics A: Mathematical and
  Theoretical}\ }\textbf {\bibinfo {volume} {51}},\ \bibinfo {pages} {375401}
  (\bibinfo {year} {2018})}\BibitemShut {NoStop}%
\bibitem [{\citenamefont {Liu}\ \emph {et~al.}(2018)\citenamefont {Liu},
  \citenamefont {Lake},\ and\ \citenamefont {Zang}}]{liu_binding_2018}%
  \BibitemOpen
  \bibfield  {author} {\bibinfo {author} {\bibfnamefont {Y.}~\bibnamefont
  {Liu}}, \bibinfo {author} {\bibfnamefont {R.~K.}\ \bibnamefont {Lake}}, \
  and\ \bibinfo {author} {\bibfnamefont {J.}~\bibnamefont {Zang}},\ }\href
  {\doibase 10.1103/PhysRevB.98.174437} {\bibfield  {journal} {\bibinfo
  {journal} {Physical Review B}\ }\textbf {\bibinfo {volume} {98}} (\bibinfo
  {year} {2018}),\ 10.1103/PhysRevB.98.174437}\BibitemShut {NoStop}%
\bibitem [{\citenamefont {Kl\'{e}man}(1985)}]{kleman_frustration_1985}%
  \BibitemOpen
  \bibfield  {author} {\bibinfo {author} {\bibfnamefont {M.}~\bibnamefont
  {Kl\'{e}man}},\ }\href {\doibase 10.1051/jphyslet:019850046016072300}
  {\bibfield  {journal} {\bibinfo  {journal} {Journal de Physique Lettres}\
  }\textbf {\bibinfo {volume} {46}},\ \bibinfo {pages} {723} (\bibinfo {year}
  {1985})}\BibitemShut {NoStop}%
\bibitem [{\citenamefont {Sadoc}\ and\ \citenamefont
  {Charvolin}(2009)}]{sadoc_3-sphere_2009}%
  \BibitemOpen
  \bibfield  {author} {\bibinfo {author} {\bibfnamefont {J.~F.}\ \bibnamefont
  {Sadoc}}\ and\ \bibinfo {author} {\bibfnamefont {J.}~\bibnamefont
  {Charvolin}},\ }\href {\doibase 10.1088/1751-8113/42/46/465209} {\bibfield
  {journal} {\bibinfo  {journal} {Journal of Physics A: Mathematical and
  Theoretical}\ }\textbf {\bibinfo {volume} {42}},\ \bibinfo {pages} {465209}
  (\bibinfo {year} {2009})}\BibitemShut {NoStop}%
\bibitem [{\citenamefont {Kuli{\'c}}\ \emph {et~al.}(2004)\citenamefont
  {Kuli{\'c}}, \citenamefont {Andrienko},\ and\ \citenamefont
  {Deserno}}]{kulic_twist-bend_2004}%
  \BibitemOpen
  \bibfield  {author} {\bibinfo {author} {\bibfnamefont {I.~M.}\ \bibnamefont
  {Kuli{\'c}}}, \bibinfo {author} {\bibfnamefont {D.}~\bibnamefont
  {Andrienko}}, \ and\ \bibinfo {author} {\bibfnamefont {M.}~\bibnamefont
  {Deserno}},\ }\href {\doibase 10.1209/epl/i2004-10076-x} {\bibfield
  {journal} {\bibinfo  {journal} {Europhysics Letters}\ }\textbf {\bibinfo
  {volume} {67}},\ \bibinfo {pages} {418} (\bibinfo {year} {2004})}\BibitemShut
  {NoStop}%
\bibitem [{\citenamefont {Grason}(2012)}]{grason_defects_2012}%
  \BibitemOpen
  \bibfield  {author} {\bibinfo {author} {\bibfnamefont {G.~M.}\ \bibnamefont
  {Grason}},\ }\href {\doibase 10.1103/PhysRevE.85.031603} {\bibfield
  {journal} {\bibinfo  {journal} {Physical Review E}\ }\textbf {\bibinfo
  {volume} {85}} (\bibinfo {year} {2012}),\
  10.1103/PhysRevE.85.031603}\BibitemShut {NoStop}%
\bibitem [{\citenamefont {Azadi}\ and\ \citenamefont
  {Grason}(2012)}]{azadi_defects_2012}%
  \BibitemOpen
  \bibfield  {author} {\bibinfo {author} {\bibfnamefont {A.}~\bibnamefont
  {Azadi}}\ and\ \bibinfo {author} {\bibfnamefont {G.~M.}\ \bibnamefont
  {Grason}},\ }\href {\doibase 10.1103/PhysRevE.85.031604} {\bibfield
  {journal} {\bibinfo  {journal} {Physical Review E}\ }\textbf {\bibinfo
  {volume} {85}} (\bibinfo {year} {2012}),\
  10.1103/PhysRevE.85.031604}\BibitemShut {NoStop}%
\bibitem [{\citenamefont {Bruss}\ and\ \citenamefont
  {Grason}(2013)}]{bruss_topological_2013}%
  \BibitemOpen
  \bibfield  {author} {\bibinfo {author} {\bibfnamefont {I.~R.}\ \bibnamefont
  {Bruss}}\ and\ \bibinfo {author} {\bibfnamefont {G.~M.}\ \bibnamefont
  {Grason}},\ }\href {\doibase 10.1039/c3sm50672j} {\bibfield  {journal}
  {\bibinfo  {journal} {Soft Matter}\ }\textbf {\bibinfo {volume} {9}},\
  \bibinfo {pages} {8327} (\bibinfo {year} {2013})}\BibitemShut {NoStop}%
\bibitem [{\citenamefont {Panaitescu}\ \emph {et~al.}(2017)\citenamefont
  {Panaitescu}, \citenamefont {Grason},\ and\ \citenamefont
  {Kudrolli}}]{panaitescu_measuring_2017}%
  \BibitemOpen
  \bibfield  {author} {\bibinfo {author} {\bibfnamefont {A.}~\bibnamefont
  {Panaitescu}}, \bibinfo {author} {\bibfnamefont {G.~M.}\ \bibnamefont
  {Grason}}, \ and\ \bibinfo {author} {\bibfnamefont {A.}~\bibnamefont
  {Kudrolli}},\ }\href {\doibase 10.1103/PhysRevE.95.052503} {\bibfield
  {journal} {\bibinfo  {journal} {Physical Review E}\ }\textbf {\bibinfo
  {volume} {95}} (\bibinfo {year} {2017}),\
  10.1103/PhysRevE.95.052503}\BibitemShut {NoStop}%
\bibitem [{\citenamefont {Panaitescu}\ \emph {et~al.}(2018)\citenamefont
  {Panaitescu}, \citenamefont {Grason},\ and\ \citenamefont
  {Kudrolli}}]{panaitescu_persistence_2018}%
  \BibitemOpen
  \bibfield  {author} {\bibinfo {author} {\bibfnamefont {A.}~\bibnamefont
  {Panaitescu}}, \bibinfo {author} {\bibfnamefont {G.~M.}\ \bibnamefont
  {Grason}}, \ and\ \bibinfo {author} {\bibfnamefont {A.}~\bibnamefont
  {Kudrolli}},\ }\href {\doibase 10.1103/PhysRevLett.120.248002} {\bibfield
  {journal} {\bibinfo  {journal} {Physical Review Letters}\ }\textbf {\bibinfo
  {volume} {120}},\ \bibinfo {pages} {248002} (\bibinfo {year}
  {2018})}\BibitemShut {NoStop}%
\bibitem [{\citenamefont {Cajamarca}\ and\ \citenamefont
  {Grason}(2014)}]{cajamarca_geometry_2014}%
  \BibitemOpen
  \bibfield  {author} {\bibinfo {author} {\bibfnamefont {L.}~\bibnamefont
  {Cajamarca}}\ and\ \bibinfo {author} {\bibfnamefont {G.~M.}\ \bibnamefont
  {Grason}},\ }\href {\doibase 10.1063/1.4900983} {\bibfield  {journal}
  {\bibinfo  {journal} {The Journal of Chemical Physics}\ }\textbf {\bibinfo
  {volume} {141}},\ \bibinfo {pages} {174904} (\bibinfo {year}
  {2014})}\BibitemShut {NoStop}%
\bibitem [{\citenamefont {Wang}\ \emph {et~al.}(2015)\citenamefont {Wang},
  \citenamefont {Ostanin}, \citenamefont {Gaidău},\ and\ \citenamefont
  {Dumitricǎ}}]{wang_twisting_2015}%
  \BibitemOpen
  \bibfield  {author} {\bibinfo {author} {\bibfnamefont {Y.}~\bibnamefont
  {Wang}}, \bibinfo {author} {\bibfnamefont {I.}~\bibnamefont {Ostanin}},
  \bibinfo {author} {\bibfnamefont {C.}~\bibnamefont {Gaidău}}, \ and\
  \bibinfo {author} {\bibfnamefont {T.}~\bibnamefont {Dumitricǎ}},\ }\href
  {\doibase 10.1021/acs.langmuir.5b03208} {\bibfield  {journal} {\bibinfo
  {journal} {Langmuir}\ }\textbf {\bibinfo {volume} {31}},\ \bibinfo {pages}
  {12323} (\bibinfo {year} {2015})}\BibitemShut {NoStop}%
\bibitem [{\citenamefont {Gonzalez}\ \emph {et~al.}(2002)\citenamefont
  {Gonzalez}, \citenamefont {Maddocks}, \citenamefont {Schuricht},\ and\
  \citenamefont {Mosel}}]{gonzalez_global_2002}%
  \BibitemOpen
  \bibfield  {author} {\bibinfo {author} {\bibfnamefont {O.}~\bibnamefont
  {Gonzalez}}, \bibinfo {author} {\bibfnamefont {J.~H.}\ \bibnamefont
  {Maddocks}}, \bibinfo {author} {\bibfnamefont {F.}~\bibnamefont {Schuricht}},
  \ and\ \bibinfo {author} {\bibfnamefont {H.~v.~d.}\ \bibnamefont {Mosel}},\
  }\href {\doibase 10.1007/s005260100089} {\bibfield  {journal} {\bibinfo
  {journal} {Calculus of Variations and Partial Differential Equations}\
  }\textbf {\bibinfo {volume} {14}},\ \bibinfo {pages} {29} (\bibinfo {year}
  {2002})}\BibitemShut {NoStop}%
\bibitem [{\citenamefont {Hall}\ \emph {et~al.}(2016)\citenamefont {Hall},
  \citenamefont {Bruss}, \citenamefont {Barone},\ and\ \citenamefont
  {Grason}}]{hall_morphology_2016}%
  \BibitemOpen
  \bibfield  {author} {\bibinfo {author} {\bibfnamefont {D.~M.}\ \bibnamefont
  {Hall}}, \bibinfo {author} {\bibfnamefont {I.~R.}\ \bibnamefont {Bruss}},
  \bibinfo {author} {\bibfnamefont {J.~R.}\ \bibnamefont {Barone}}, \ and\
  \bibinfo {author} {\bibfnamefont {G.~M.}\ \bibnamefont {Grason}},\ }\href
  {\doibase 10.1038/nmat4598} {\bibfield  {journal} {\bibinfo  {journal}
  {Nature Materials}\ }\textbf {\bibinfo {volume} {15}},\ \bibinfo {pages}
  {727} (\bibinfo {year} {2016})}\BibitemShut {NoStop}%
\bibitem [{Note1()}]{Note1}%
  \BibitemOpen
  \bibinfo {note} {In general, there are may be multiple extrema of ${\setbox
  \z@ \hbox {\frozen@everymath \@emptytoks \mathsurround \z@
  $\nulldelimiterspace \z@ \left |\vcenter to\@ne \big@size {}\right .$}\box
  \z@ }\protect \mathbf {r}_1(s_1) - \protect \mathbf {r}_2(s_2){\setbox \z@
  \hbox {\frozen@everymath \@emptytoks \mathsurround \z@ $\nulldelimiterspace
  \z@ \left |\vcenter to\@ne \big@size {}\right .$}\box \z@ }^2$, corresponding
  to multiple solutions for $s_2(s_1)$ to Eq. (\ref {closestapproacheqn}), for
  a given pair. Our analysis assumes the minimal distance for solutions
  $s_2(s_1)$ for a given $s_1$.}\BibitemShut {Stop}%
\bibitem [{\citenamefont {Cantarella}\ \emph {et~al.}(2002)\citenamefont
  {Cantarella}, \citenamefont {Kusner},\ and\ \citenamefont
  {Sullivan}}]{cantarella_minimum_2002}%
  \BibitemOpen
  \bibfield  {author} {\bibinfo {author} {\bibfnamefont {J.}~\bibnamefont
  {Cantarella}}, \bibinfo {author} {\bibfnamefont {R.~B.}\ \bibnamefont
  {Kusner}}, \ and\ \bibinfo {author} {\bibfnamefont {J.~M.}\ \bibnamefont
  {Sullivan}},\ }\href {\doibase 10.1007/s00222-002-0234-y} {\bibfield
  {journal} {\bibinfo  {journal} {Inventiones mathematicae}\ }\textbf {\bibinfo
  {volume} {150}},\ \bibinfo {pages} {257} (\bibinfo {year}
  {2002})}\BibitemShut {NoStop}%
\bibitem [{\citenamefont {Gonzalez}\ and\ \citenamefont
  {Maddocks}(1999)}]{gonzalez_global_1999}%
  \BibitemOpen
  \bibfield  {author} {\bibinfo {author} {\bibfnamefont {O.}~\bibnamefont
  {Gonzalez}}\ and\ \bibinfo {author} {\bibfnamefont {J.~H.}\ \bibnamefont
  {Maddocks}},\ }\href {\doibase 10.1073/pnas.96.9.4769} {\bibfield  {journal}
  {\bibinfo  {journal} {Proceedings of the National Academy of Sciences}\
  }\textbf {\bibinfo {volume} {96}},\ \bibinfo {pages} {4769} (\bibinfo {year}
  {1999})}\BibitemShut {NoStop}%
\bibitem [{Note2()}]{Note2}%
  \BibitemOpen
  \bibinfo {note} {We again adopt the reparameterization of $\protect \mathbf
  {r}_2$ in terms of the arc length of $\protect \mathbf {r}_1$, which we call
  $s$ for simplicity of notation.}\BibitemShut {Stop}%
\bibitem [{\citenamefont {Achard}\ \emph {et~al.}(2005)\citenamefont {Achard},
  \citenamefont {Kl{\'e}man}, \citenamefont {Nastishin},\ and\ \citenamefont
  {Nguyen}}]{achard_liquid_2005}%
  \BibitemOpen
  \bibfield  {author} {\bibinfo {author} {\bibfnamefont {M.-F.}\ \bibnamefont
  {Achard}}, \bibinfo {author} {\bibfnamefont {M.}~\bibnamefont {Kl{\'e}man}},
  \bibinfo {author} {\bibfnamefont {Y.~A.}\ \bibnamefont {Nastishin}}, \ and\
  \bibinfo {author} {\bibfnamefont {H.-T.}\ \bibnamefont {Nguyen}},\ }\href
  {\doibase 10.1140/epje/e2005-00005-2} {\bibfield  {journal} {\bibinfo
  {journal} {The European Physical Journal E}\ }\textbf {\bibinfo {volume}
  {16}},\ \bibinfo {pages} {37} (\bibinfo {year} {2005})}\BibitemShut {NoStop}%
\bibitem [{\citenamefont {Coleman}\ \emph {et~al.}(2003)\citenamefont
  {Coleman}, \citenamefont {Fernsler}, \citenamefont {Chattham}, \citenamefont
  {Nakata}, \citenamefont {Takanishi}, \citenamefont {Körblova}, \citenamefont
  {Link}, \citenamefont {Shao}, \citenamefont {Jang},\ and\ \citenamefont
  {Maclennan}}]{coleman_polarization-modulated_2003}%
  \BibitemOpen
  \bibfield  {author} {\bibinfo {author} {\bibfnamefont {D.~A.}\ \bibnamefont
  {Coleman}}, \bibinfo {author} {\bibfnamefont {J.}~\bibnamefont {Fernsler}},
  \bibinfo {author} {\bibfnamefont {N.}~\bibnamefont {Chattham}}, \bibinfo
  {author} {\bibfnamefont {M.}~\bibnamefont {Nakata}}, \bibinfo {author}
  {\bibfnamefont {Y.}~\bibnamefont {Takanishi}}, \bibinfo {author}
  {\bibfnamefont {E.}~\bibnamefont {Körblova}}, \bibinfo {author}
  {\bibfnamefont {D.~R.}\ \bibnamefont {Link}}, \bibinfo {author}
  {\bibfnamefont {R.-F.}\ \bibnamefont {Shao}}, \bibinfo {author}
  {\bibfnamefont {W.~G.}\ \bibnamefont {Jang}}, \ and\ \bibinfo {author}
  {\bibfnamefont {J.~E.}\ \bibnamefont {Maclennan}},\ }\href@noop {} {\bibfield
   {journal} {\bibinfo  {journal} {Science}\ }\textbf {\bibinfo {volume}
  {301}},\ \bibinfo {pages} {1204} (\bibinfo {year} {2003})}\BibitemShut
  {NoStop}%
\bibitem [{\citenamefont {Grason}(2015)}]{grason_colloquium_2015}%
  \BibitemOpen
  \bibfield  {author} {\bibinfo {author} {\bibfnamefont {G.~M.}\ \bibnamefont
  {Grason}},\ }\href {\doibase 10.1103/RevModPhys.87.401} {\bibfield  {journal}
  {\bibinfo  {journal} {Reviews of Modern Physics}\ }\textbf {\bibinfo {volume}
  {87}},\ \bibinfo {pages} {401} (\bibinfo {year} {2015})}\BibitemShut
  {NoStop}%
\bibitem [{\citenamefont {Wright}\ and\ \citenamefont
  {Mermin}(1989)}]{wright_crystalline_1989}%
  \BibitemOpen
  \bibfield  {author} {\bibinfo {author} {\bibfnamefont {D.~C.}\ \bibnamefont
  {Wright}}\ and\ \bibinfo {author} {\bibfnamefont {N.~D.}\ \bibnamefont
  {Mermin}},\ }\href@noop {} {\bibfield  {journal} {\bibinfo  {journal}
  {Reviews of Modern Physics}\ }\textbf {\bibinfo {volume} {61}},\ \bibinfo
  {pages} {385} (\bibinfo {year} {1989})}\BibitemShut {NoStop}%
\bibitem [{\citenamefont {Sethna}\ \emph {et~al.}(1983)\citenamefont {Sethna},
  \citenamefont {Wright},\ and\ \citenamefont
  {Mermin}}]{sethna_relieving_1983}%
  \BibitemOpen
  \bibfield  {author} {\bibinfo {author} {\bibfnamefont {J.~P.}\ \bibnamefont
  {Sethna}}, \bibinfo {author} {\bibfnamefont {D.~C.}\ \bibnamefont {Wright}},
  \ and\ \bibinfo {author} {\bibfnamefont {N.~D.}\ \bibnamefont {Mermin}},\
  }\href@noop {} {\bibfield  {journal} {\bibinfo  {journal} {Physical Review
  Letters}\ }\textbf {\bibinfo {volume} {51}},\ \bibinfo {pages} {467}
  (\bibinfo {year} {1983})}\BibitemShut {NoStop}%
\bibitem [{\citenamefont {Gromoll}\ and\ \citenamefont
  {Walschap}(2009)}]{gromoll_metric_2009}%
  \BibitemOpen
  \bibfield  {author} {\bibinfo {author} {\bibfnamefont {D.}~\bibnamefont
  {Gromoll}}\ and\ \bibinfo {author} {\bibfnamefont {G.}~\bibnamefont
  {Walschap}},\ }\enquote {\bibinfo {title} {Submersions, foliations, and
  metrics},}\ in\ \href {\doibase 10.1007/978-3-7643-8715-0_1} {\emph {\bibinfo
  {booktitle} {Metric Foliations and Curvature}}}\ (\bibinfo  {publisher}
  {Birkh{\"a}user Basel},\ \bibinfo {address} {Basel},\ \bibinfo {year}
  {2009})\ pp.\ \bibinfo {pages} {1--44}\BibitemShut {NoStop}%
\bibitem [{\citenamefont {Rey}(2010)}]{rey_liquid_2010}%
  \BibitemOpen
  \bibfield  {author} {\bibinfo {author} {\bibfnamefont {A.~D.}\ \bibnamefont
  {Rey}},\ }\href {\doibase 10.1039/B921576J} {\bibfield  {journal} {\bibinfo
  {journal} {Soft Matter}\ }\textbf {\bibinfo {volume} {6}},\ \bibinfo {pages}
  {3402} (\bibinfo {year} {2010})}\BibitemShut {NoStop}%
\bibitem [{\citenamefont {Shin}\ and\ \citenamefont
  {Grason}(2011)}]{shin_filling_2011}%
  \BibitemOpen
  \bibfield  {author} {\bibinfo {author} {\bibfnamefont {H.}~\bibnamefont
  {Shin}}\ and\ \bibinfo {author} {\bibfnamefont {G.~M.}\ \bibnamefont
  {Grason}},\ }\href {\doibase 10.1209/0295-5075/96/36007} {\bibfield
  {journal} {\bibinfo  {journal} {Europhysics Letters}\ }\textbf {\bibinfo
  {volume} {96}},\ \bibinfo {pages} {36007} (\bibinfo {year}
  {2011})}\BibitemShut {NoStop}%
\bibitem [{\citenamefont {Brown}\ \emph {et~al.}(2014)\citenamefont {Brown},
  \citenamefont {Kreplak},\ and\ \citenamefont
  {Rutenberg}}]{brown_equilibrium_2014}%
  \BibitemOpen
  \bibfield  {author} {\bibinfo {author} {\bibfnamefont {A.~I.}\ \bibnamefont
  {Brown}}, \bibinfo {author} {\bibfnamefont {L.}~\bibnamefont {Kreplak}}, \
  and\ \bibinfo {author} {\bibfnamefont {A.~D.}\ \bibnamefont {Rutenberg}},\
  }\href {\doibase 10.1039/C4SM01359J} {\bibfield  {journal} {\bibinfo
  {journal} {Soft Matter}\ }\textbf {\bibinfo {volume} {10}},\ \bibinfo {pages}
  {8500} (\bibinfo {year} {2014})}\BibitemShut {NoStop}%
\bibitem [{\citenamefont {de~Gennes}\ and\ \citenamefont
  {Prost}(1995)}]{degennes_physics_1995}%
  \BibitemOpen
  \bibfield  {author} {\bibinfo {author} {\bibfnamefont {P.~G.}\ \bibnamefont
  {de~Gennes}}\ and\ \bibinfo {author} {\bibfnamefont {J.}~\bibnamefont
  {Prost}},\ }\href@noop {} {\emph {\bibinfo {title} {The {Physics} of {Liquid}
  {Crystals}}}},\ \bibinfo {edition} {second edition}\ ed.,\ International
  {Series} of {Monographs} on {Physics}\ (\bibinfo  {publisher} {Oxford
  University Press},\ \bibinfo {address} {Oxford, New York},\ \bibinfo {year}
  {1995})\BibitemShut {NoStop}%
\bibitem [{\citenamefont {Grason}(2009)}]{grason_braided_2009}%
  \BibitemOpen
  \bibfield  {author} {\bibinfo {author} {\bibfnamefont {G.~M.}\ \bibnamefont
  {Grason}},\ }\href {\doibase 10.1103/PhysRevE.79.041919} {\bibfield
  {journal} {\bibinfo  {journal} {Physical Review E}\ }\textbf {\bibinfo
  {volume} {79}} (\bibinfo {year} {2009}),\
  10.1103/PhysRevE.79.041919}\BibitemShut {NoStop}%
\bibitem [{\citenamefont {Millman}\ and\ \citenamefont
  {Parker}(1977)}]{millman_elements_1977}%
  \BibitemOpen
  \bibfield  {author} {\bibinfo {author} {\bibfnamefont {R.~S.}\ \bibnamefont
  {Millman}}\ and\ \bibinfo {author} {\bibfnamefont {G.~D.}\ \bibnamefont
  {Parker}},\ }\href@noop {} {\emph {\bibinfo {title} {Elements of Differential
  Geometry}}}\ (\bibinfo  {publisher} {Englewood Cliffs, N.J. :
  Prentice-Hall},\ \bibinfo {year} {1977})\BibitemShut {NoStop}%
\bibitem [{\citenamefont {Neville}(1993)}]{neville_biology_1993}%
  \BibitemOpen
  \bibfield  {author} {\bibinfo {author} {\bibfnamefont {A.~C.}\ \bibnamefont
  {Neville}},\ }\href {\doibase 10.1017/CBO9780511601101} {\emph {\bibinfo
  {title} {Biology of Fibrous Composites: Development beyond the Cell
  Membrane}}}\ (\bibinfo  {publisher} {Cambridge University Press},\ \bibinfo
  {year} {1993})\BibitemShut {NoStop}%
\bibitem [{\citenamefont {Bouligand}(2008)}]{bouligand_liquid_2008}%
  \BibitemOpen
  \bibfield  {author} {\bibinfo {author} {\bibfnamefont {Y.}~\bibnamefont
  {Bouligand}},\ }\href {\doibase 10.1016/j.crci.2007.10.001} {\bibfield
  {journal} {\bibinfo  {journal} {Comptes Rendus Chimie}\ }\textbf {\bibinfo
  {volume} {11}},\ \bibinfo {pages} {281} (\bibinfo {year} {2008})}\BibitemShut
  {NoStop}%
\bibitem [{\citenamefont {Hall}\ and\ \citenamefont
  {Grason}(2017)}]{hall_how_2017}%
  \BibitemOpen
  \bibfield  {author} {\bibinfo {author} {\bibfnamefont {D.~M.}\ \bibnamefont
  {Hall}}\ and\ \bibinfo {author} {\bibfnamefont {G.~M.}\ \bibnamefont
  {Grason}},\ }\href {\doibase 10.1098/rsfs.2016.0140} {\bibfield  {journal}
  {\bibinfo  {journal} {Interface Focus}\ }\textbf {\bibinfo {volume} {7}},\
  \bibinfo {pages} {20160140} (\bibinfo {year} {2017})}\BibitemShut {NoStop}%
\bibitem [{\citenamefont {Koning}\ \emph {et~al.}(2014)\citenamefont {Koning},
  \citenamefont {Zuiden}, \citenamefont {Kamien},\ and\ \citenamefont
  {Vitelli}}]{koning_saddle-splay_2014}%
  \BibitemOpen
  \bibfield  {author} {\bibinfo {author} {\bibfnamefont {V.}~\bibnamefont
  {Koning}}, \bibinfo {author} {\bibfnamefont {B.~C.~v.}\ \bibnamefont
  {Zuiden}}, \bibinfo {author} {\bibfnamefont {R.~D.}\ \bibnamefont {Kamien}},
  \ and\ \bibinfo {author} {\bibfnamefont {V.}~\bibnamefont {Vitelli}},\ }\href
  {\doibase 10.1039/C4SM00076E} {\bibfield  {journal} {\bibinfo  {journal}
  {Soft Matter}\ }\textbf {\bibinfo {volume} {10}},\ \bibinfo {pages} {4192}
  (\bibinfo {year} {2014})}\BibitemShut {NoStop}%
\bibitem [{\citenamefont {Sadoc}\ and\ \citenamefont
  {Mosseri}(1999)}]{sadoc_geometrical_1999}%
  \BibitemOpen
  \bibfield  {author} {\bibinfo {author} {\bibfnamefont {J.-F.}\ \bibnamefont
  {Sadoc}}\ and\ \bibinfo {author} {\bibfnamefont {R.}~\bibnamefont
  {Mosseri}},\ }\href {\doibase 10.1017/CBO9780511599934} {\emph {\bibinfo
  {title} {Geometrical Frustration}}},\ Collection Alea-Saclay: Monographs and
  Texts in Statistical Physics\ (\bibinfo  {publisher} {Cambridge University
  Press},\ \bibinfo {year} {1999})\BibitemShut {NoStop}%
\bibitem [{\citenamefont {Charvolin}\ and\ \citenamefont
  {Sadoc}(2008)}]{charvolin_geometrical_2008}%
  \BibitemOpen
  \bibfield  {author} {\bibinfo {author} {\bibfnamefont {J.}~\bibnamefont
  {Charvolin}}\ and\ \bibinfo {author} {\bibfnamefont {J.~F.}\ \bibnamefont
  {Sadoc}},\ }\href {\doibase 10.1140/epje/i2008-10313-8} {\bibfield  {journal}
  {\bibinfo  {journal} {The European Physical Journal E}\ }\textbf {\bibinfo
  {volume} {25}},\ \bibinfo {pages} {335} (\bibinfo {year} {2008})}\BibitemShut
  {NoStop}%
\bibitem [{\citenamefont {Mosseri}\ and\ \citenamefont
  {Sadoc}(2012)}]{mosseri_hopf_2012}%
  \BibitemOpen
  \bibfield  {author} {\bibinfo {author} {\bibfnamefont {R.}~\bibnamefont
  {Mosseri}}\ and\ \bibinfo {author} {\bibfnamefont {J.-F.}\ \bibnamefont
  {Sadoc}},\ }\href {\doibase 10.1007/s11224-012-0010-6} {\bibfield  {journal}
  {\bibinfo  {journal} {Structural Chemistry}\ }\textbf {\bibinfo {volume}
  {23}},\ \bibinfo {pages} {1071} (\bibinfo {year} {2012})}\BibitemShut
  {NoStop}%
\bibitem [{Note3()}]{Note3}%
  \BibitemOpen
  \bibinfo {note} {Strictly, $\alpha $ is a rational number such that the
  $(a,b)$ Seifert fibration of $S^{3}$ has $\alpha = \protect \genfrac
  {}{}{}1{a}{b}$.}\BibitemShut {Stop}%
\bibitem [{\citenamefont {de~Gennes}(1976)}]{de_gennes_polymeric_1976}%
  \BibitemOpen
  \bibfield  {author} {\bibinfo {author} {\bibfnamefont {P.~G.}\ \bibnamefont
  {de~Gennes}},\ }\href {\doibase 10.1080/15421407708083702} {\bibfield
  {journal} {\bibinfo  {journal} {Molecular Crystals and Liquid Crystals}\
  }\textbf {\bibinfo {volume} {34}},\ \bibinfo {pages} {177} (\bibinfo {year}
  {1976})}\BibitemShut {NoStop}%
\bibitem [{Note4()}]{Note4}%
  \BibitemOpen
  \bibinfo {note} {It can be shown that neglect of the constant circulation
  constraint of Eq.~\protect \textup {\hbox {\mathsurround \z@ \protect
  \normalfont (\ignorespaces \ref {eq:trfree_pitch}\unskip \@@italiccorr )}} in
  the linear ansatz $f(\rho ) = \Omega \rho $ studied in refs.~\cite
  {kulic_twist-bend_2004,koning_saddle-splay_2014} leads to less equidistanct
  splay-free textures, with $\delta r \sim {\protect \rm min}[\Omega , \kappa
  _0] \Omega \rho ^2$}\BibitemShut {NoStop}%
\bibitem [{\citenamefont {Brodsky}\ and\ \citenamefont
  {Persikov}(2005)}]{brodsky_molecular_2005}%
  \BibitemOpen
  \bibfield  {author} {\bibinfo {author} {\bibfnamefont {B.}~\bibnamefont
  {Brodsky}}\ and\ \bibinfo {author} {\bibfnamefont {A.~V.}\ \bibnamefont
  {Persikov}},\ }in\ \href {\doibase 10.1016/S0065-3233(05)70009-7} {\emph
  {\bibinfo {booktitle} {Advances in {Protein} {Chemistry}}}},\ \bibinfo
  {series} {Fibrous {Proteins}: {Coiled}-{Coils}, {Collagen} and {Elastomers}},
  Vol.~\bibinfo {volume} {70}\ (\bibinfo  {publisher} {Academic Press},\
  \bibinfo {year} {2005})\ pp.\ \bibinfo {pages} {301--339}\BibitemShut
  {NoStop}%
\bibitem [{\citenamefont {Katritch}\ \emph {et~al.}(1996)\citenamefont
  {Katritch}, \citenamefont {Bednar}, \citenamefont {Michoud}, \citenamefont
  {Scharein}, \citenamefont {Dubochet},\ and\ \citenamefont
  {Stasiak}}]{katritch_geometry_1996}%
  \BibitemOpen
  \bibfield  {author} {\bibinfo {author} {\bibfnamefont {V.}~\bibnamefont
  {Katritch}}, \bibinfo {author} {\bibfnamefont {J.}~\bibnamefont {Bednar}},
  \bibinfo {author} {\bibfnamefont {D.}~\bibnamefont {Michoud}}, \bibinfo
  {author} {\bibfnamefont {R.~G.}\ \bibnamefont {Scharein}}, \bibinfo {author}
  {\bibfnamefont {J.}~\bibnamefont {Dubochet}}, \ and\ \bibinfo {author}
  {\bibfnamefont {A.}~\bibnamefont {Stasiak}},\ }\href {\doibase
  10.1038/384142a0} {\bibfield  {journal} {\bibinfo  {journal} {Nature}\
  }\textbf {\bibinfo {volume} {384}},\ \bibinfo {pages} {142} (\bibinfo {year}
  {1996})}\BibitemShut {NoStop}%
\bibitem [{\citenamefont {Carlen}\ \emph {et~al.}(2005)\citenamefont {Carlen},
  \citenamefont {Laurie}, \citenamefont {Maddocks},\ and\ \citenamefont
  {Smutny}}]{calvo_biarcs_2005}%
  \BibitemOpen
  \bibfield  {author} {\bibinfo {author} {\bibfnamefont {M.}~\bibnamefont
  {Carlen}}, \bibinfo {author} {\bibfnamefont {B.}~\bibnamefont {Laurie}},
  \bibinfo {author} {\bibfnamefont {J.~H.}\ \bibnamefont {Maddocks}}, \ and\
  \bibinfo {author} {\bibfnamefont {J.}~\bibnamefont {Smutny}},\ }in\ \href
  {\doibase 10.1142/9789812703460_0005} {\emph {\bibinfo {booktitle} {Physical
  and Numerical Models in Knot Theory}}},\ \bibinfo {series} {Series on {Knots}
  and {Everything}}, Vol.~\bibinfo {volume} {36}\ (\bibinfo  {publisher} {WORLD
  SCIENTIFIC},\ \bibinfo {year} {2005})\ Chap.~\bibinfo {chapter} {5}, pp.\
  \bibinfo {pages} {75--108}\BibitemShut {NoStop}%
\bibitem [{\citenamefont {Calvo}\ \emph {et~al.}(2005)\citenamefont {Calvo},
  \citenamefont {Millett}, \citenamefont {Rawdon},\ and\ \citenamefont
  {Stasiak}}]{calvo_physical_2005}%
  \BibitemOpen
  \bibfield  {author} {\bibinfo {author} {\bibfnamefont {J.~A.}\ \bibnamefont
  {Calvo}}, \bibinfo {author} {\bibfnamefont {K.~C.}\ \bibnamefont {Millett}},
  \bibinfo {author} {\bibfnamefont {E.~J.}\ \bibnamefont {Rawdon}}, \ and\
  \bibinfo {author} {\bibfnamefont {A.}~\bibnamefont {Stasiak}},\ }\href
  {\doibase 10.1142/5766} {\emph {\bibinfo {title} {Physical and Numerical
  Models in Knot Theory}}},\ \bibinfo {series} {Series on {Knots} and
  {Everything}}, Vol.~\bibinfo {volume} {36}\ (\bibinfo  {publisher} {WORLD
  SCIENTIFIC},\ \bibinfo {year} {2005})\BibitemShut {NoStop}%
\bibitem [{\citenamefont {Janse~van
  Rensburg}(2005)}]{janse_van_rensburg_tutorial_2005}%
  \BibitemOpen
  \bibfield  {author} {\bibinfo {author} {\bibfnamefont {E.~J.}\ \bibnamefont
  {Janse~van Rensburg}},\ }in\ \href {\doibase 10.1142/9789812703460_0002}
  {\emph {\bibinfo {booktitle} {Physical and {Numerical} {Models} in {Knot}
  {Theory}}}},\ \bibinfo {series} {Series on {Knots} and {Everything}},
  Vol.~\bibinfo {volume} {36}\ (\bibinfo  {publisher} {WORLD SCIENTIFIC},\
  \bibinfo {year} {2005})\ Chap.~\bibinfo {chapter} {2}, pp.\ \bibinfo {pages}
  {19--44}\BibitemShut {NoStop}%
\bibitem [{\citenamefont {Faddeev}\ and\ \citenamefont
  {Niemi}(1997)}]{faddeev_stable_1997}%
  \BibitemOpen
  \bibfield  {author} {\bibinfo {author} {\bibfnamefont {L.}~\bibnamefont
  {Faddeev}}\ and\ \bibinfo {author} {\bibfnamefont {A.~J.}\ \bibnamefont
  {Niemi}},\ }\href {\doibase 10.1038/387058a0} {\bibfield  {journal} {\bibinfo
   {journal} {Nature}\ }\textbf {\bibinfo {volume} {387}},\ \bibinfo {pages}
  {58} (\bibinfo {year} {1997})}\BibitemShut {NoStop}%
\bibitem [{\citenamefont {Battye}\ and\ \citenamefont
  {Sutcliffe}(1998)}]{battye_knots_1998}%
  \BibitemOpen
  \bibfield  {author} {\bibinfo {author} {\bibfnamefont {R.~A.}\ \bibnamefont
  {Battye}}\ and\ \bibinfo {author} {\bibfnamefont {P.~M.}\ \bibnamefont
  {Sutcliffe}},\ }\href {\doibase 10.1103/PhysRevLett.81.4798} {\bibfield
  {journal} {\bibinfo  {journal} {Physical Review Letters}\ }\textbf {\bibinfo
  {volume} {81}},\ \bibinfo {pages} {4798} (\bibinfo {year}
  {1998})}\BibitemShut {NoStop}%
\bibitem [{\citenamefont {Bruss}\ and\ \citenamefont
  {Grason}(2018)}]{bruss_defect-driven_2018}%
  \BibitemOpen
  \bibfield  {author} {\bibinfo {author} {\bibfnamefont {I.~R.}\ \bibnamefont
  {Bruss}}\ and\ \bibinfo {author} {\bibfnamefont {G.~M.}\ \bibnamefont
  {Grason}},\ }\href {\doibase 10.1103/PhysRevX.8.031046} {\bibfield  {journal}
  {\bibinfo  {journal} {Physical Review X}\ }\textbf {\bibinfo {volume} {8}},\
  \bibinfo {pages} {031046} (\bibinfo {year} {2018})}\BibitemShut {NoStop}%
\end{thebibliography}%

\end{document}